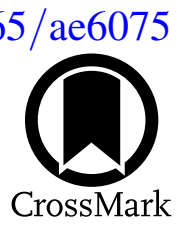

# Characterizing the Origins of the Gamma-Ray Variability of the Jetted Active Galactic Nuclei Observed with Fermi-LAT

Yongyun Chen (陈永云)[1], Qiusheng Gu (顾秋生)[2], Junhui Fan (樊军辉)[3], Dingrong Xiong (熊定荣)[4], Xiaoling Yu (俞效龄)[1], Xiaogu Zhong (钟晓谷)[1], Xiaotong Guo (郭晓通)[5], and Nan Ding (丁楠)[6]
[1] College of Physics and Electronic Engineering, Qujing Normal University, Qujing 655011, People's Republic of China; ynkmcyy@yeah
[2] School of Astronomy and Space Science, Nanjing University, Nanjing 210093, People's Republic of China; qsgu@nju.edu.cn
[3] Center for Astrophysics, Guang zhou University, Guangzhou 510006, People's Republic of China
[4] Yunnan Observatories, Chinese Academy of Sciences, Kunming 650011, People's Republic of China; xiongdingrong@ynao.ac.cn
[5] Anqing Normal University, 246133, People's Republic of China
[6] School of Physical Science and Technology, Kunming University, 650214, People's Republic of China

## Abstract

We have conducted an analysis of gamma-ray variability on a large sample of jetted active galactic nuclei (AGNs) by utilizing archival Fermi Large Area Telescope light curves and applying damped random walk modeling to obtain the variability amplitude. Our primary findings are summarized as follows. (1) The mean variability damping timescales of our sources are approximately 100 days. This damping timescale may imply that diffusive shock acceleration plays an important role in the variability of gamma-ray emission. (2) Flat-spectrum radio quasars demonstrate greater variability amplitude compared to BL Lacertae objects. (3) The ratio of the distance of the emission region from the central supermassive black hole to the dusty torus radius for our sources is $R \approx 2$–$4.5R_{\rm DT}$. In contrast, the ratio of the distance of the emission region from the central supermassive black hole to the broad-line region (BLR) radius for our sources is $R \approx 135$–$295R_{\rm BLR-in}$ and $R \approx 123$–$270R_{\rm BLR-out}$. These findings indicate that the gamma-ray emission region in jetted AGNs is likely located beyond the BLR and potentially could be associated with the dusty torus. (4) A statistical correlation is observed between the variability amplitude and radio luminosity, radio loudness, X-ray luminosity, X-ray loudness, gamma-ray luminosity, and gamma-ray loudness, indicating a potential relationship between gamma-ray variability and jet activity. (5) The variability amplitude also shows a statistical correlation with the synchrotron peak frequency luminosity, inverse-Compton peak frequency luminosity, and Compton dominance. (6) The variability amplitude also correlates with the black hole mass, accretion-disk luminosity, and Eddington ratio, implying that the accretion disk may also contribute to gamma-ray variability.

*Unified Astronomy Thesaurus concepts:* Active galactic nuclei (16); Gamma-rays (637); Jets (870); Light curves (918); Time series analysis (1916)

## 1. Introduction

Variability represents a critical characteristic of active galactic nuclei (AGNs), encompassing nonperiodic flux variations across different amplitudes and timescales, thereby serving as a fundamental tool for probing the AGN central engine (P. Padovani et al. 2017). Variability analyses hold particular significance in gamma-ray astronomy for several reasons. First, they facilitate the detection of weak sources and enable the differentiation of genuine point sources from background fluctuations. Second, the analysis of multiwavelength variability aids in identifying the correct radio, optical, or X-ray source counterpart within the positional uncertainty region of a gamma-ray source (A. A. Abdo et al. 2010a). Furthermore, characterizing individual gamma-ray variability from unidentified sources can assist in determining the appropriate source classification (P. L. Nolan et al. 2003). Concurrently, gamma-ray variability is a potent and direct diagnostic for investigating the complex physical mechanisms operating in the innermost regions of AGN jets.

The variability of AGNs has been conclusively detected across the entire electromagnetic spectrum, with associated timescales ranging from several decades down to just a few minutes (e.g., F. Aharonian et al. 2007; J. Albert et al. 2007; T. Arlen et al. 2013; J. Aleksić et al. 2014; M. Ackermann et al. 2016; A. Shukla et al. 2018; A. Sharma et al. 2026). Radio-loud AGNs display notable variability in their gamma-ray emission. Blazars represent an extreme subclass of radio-loud AGNs, characterized by relativistic jets oriented directly along the line of sight to the observer (C. M. Urry & P. Padovani 1995). These objects are strong emitters of high-energy radiation and exhibit rapid and pronounced flux variations across the entire observable electromagnetic spectrum, spanning from low-energy radio waves to high-energy gamma rays (M.-H. Ulrich et al. 1997). Blazars constitute the dominant population of extragalactic sources in the gamma-ray sky, as evidenced by observations from the Compton Gamma Ray Observatory (R. C. Hartman et al. 1999) and, more recently, the Fermi Gamma-ray Space Telescope (S. Abdollahi et al. 2020). Blazars are further classified into two main categories—BL Lacertae objects (BL Lacs) and flat-spectrum radio quasars (FSRQs)—based on the strength of their optical emission lines (e.g., M. Ajello et al. 2020). The phenomenon of gamma-ray variability is not exclusive to blazars, which are well known for their prominent and persistent flux changes, but has also been observed in misaligned jet systems, such as radio galaxies (MAGIC Collaboration et al. 2018; F. Ait Benkhali et al. 2019).

Blazars have been extensively investigated for their flux variations across multiple wavelength regimes, including the radio, optical, infrared, and X-ray bands. Among the various theoretical models proposed to account for the observed variability in blazars, the shock-in-jet model holds a prominent place. Initially introduced by A. P. Marscher & W. K. Gear (1985), this model has since been further developed and refined, particularly by M. Böttcher & C. D. Dermer (2010). In addition to the shock-in-jet model, alternative explanations for blazar variability have been proposed, such as the jet–star interaction model (M. V. Barkov et al. 2012) and magnetic reconnection processes (D. Giannios 2013). Although these models offer plausible interpretations for the observed flux variations, the precise physical mechanisms responsible for such variability remain incompletely understood. Consequently, additional research is necessary to elucidate the underlying physical processes and enhance our comprehension of blazar flux variability. The gamma-ray band remains one of the least studied regions of the electromagnetic spectrum in terms of flux variability—a situation largely due to the limited availability of measurements for a sufficiently large sample of sources.

Studies of gamma-ray variability have become a fundamental tool for probing the physical mechanisms in relativistic jets originating from AGNs and also microquasars. Since the launch of the Fermi Gamma-ray Space Telescope in 2008, numerous studies have been conducted on the gamma-ray variability of blazars (E. W. Bonning et al. 2009; A. A. Abdo et al. 2010a; R. Chatterjee et al. 2012; V. S. Paliya et al. 2015; S. Li et al. 2018; Z. Shah et al. 2018; M. Meyer et al. 2019; B. Rajput et al. 2019, 2020; G. Bhatta & N. Dhital 2020; H. Zhang et al. 2022, 2023, 2024). A. A. Abdo et al. (2010a) systematically analyzed the gamma-ray variability of 106 AGNs based on the first 11 months of Fermi survey data. Their findings indicated that more than 50% of the sources exhibited variability with a power-law (PL) power spectral density (PSD), suggesting the presence of an underlying random walk mechanism in certain blazars. K. Nakagawa & M. Mori (2013) identified a characteristic timescale of approximately 7.9 days in the PSD of 3C 454.3 through the analysis of 4 yr of Fermi Large Area Telescope (LAT) data and proposed an internal shock model to account for the origin of this timescale. M. Meyer et al. (2019) performed a comprehensive analysis of Fermi-LAT light curves for bright gamma-ray FSRQs, yielding important constraints on the physical mechanisms responsible for blazar jet emission. G. Bhatta & N. Dhital (2020) carried out an extensive and systematic variability study of a sample of 20 powerful blazars. The study revealed that blazars with steeper gamma-ray spectral indices tend to exhibit higher levels of variability, and that the distribution of the gamma-ray flux closely follows a log-normal probability distribution function.

A stochastic process model has been widely utilized to describe the optical variability of AGN accretion disks (e.g., B. C. Kelly et al. 2009; C. L. MacLeod et al. 2010; Y. Zu et al. 2013; S. Rakshit & C. S. Stalin 2017; S. Li et al. 2018; H. Zhang et al. 2018, 2022, 2023, 2024; K.-X. Lu et al. 2019; C. J. Burke et al. 2021; Z. Stone et al. 2022; D. Xiong et al. 2025). This approach has demonstrated significant effectiveness in analyzing variability across multiple wavelengths in AGN emissions. The underlying random behavior is commonly modeled using the damped random walk (DRW), equivalent to the Ornstein–Uhlenbeck process, which exhibits a characteristic break in its PSD. Typically, the PSDs associated with jets and accretion disks are fitted using a bending PL form. At frequencies higher than the break frequency—corresponding to a characteristic timescale—the PSD declines as a PL with a slope of approximately 2. In contrast, below the break frequency, the PSD flattens into a white-noise regime. This stochastic framework successfully describes the long-term variability observed in AGN accretion disks. Some authors have suggested that the stochastic process offers a robust framework for extracting meaningful insights from AGN variability (V. P. Kasliwal et al. 2017; C. J. Burke et al. 2021). In recent years, the DRW model has been extensively utilized not only in the optical time domain, but also in the submillimeter and for the gamma-ray variability in AGNs (M. A. Sobolewska et al. 2014; A. Goyal et al. 2018; J. L. Ryan et al. 2019; S. Covino et al. 2020; M. Tarnopolski et al. 2020; S. Yang et al. 2021; H. Zhang et al. 2021, 2022, 2023, 2024). The DRW model incorporates celerite—a recently developed method for modeling light curves within a stochastic process framework (D. Foreman-Mackey et al. 2017). This approach has been applied to Fermi-LAT data from AGNs, to evaluate the statistical significance of gamma-ray quasiperiodic oscillations (S. Yang et al. 2021; H. Zhang et al. 2021).

Although many authors have studied the gamma-ray variability of AGNs, they are based on small samples or individual sources and cannot obtain robust statistical results. At the same time, no author has used the DRW model to obtain the amplitude and timescales of gamma-ray variability of large samples of AGNs to study the origin of gamma-ray variability. At present, the origin of gamma-ray variability is unclear. Therefore, we use a large sample of AGNs to study the origin of gamma-ray variability. The number of blazars detected as gamma-ray emitters has substantially increased since the first such discovery. In addition, gamma-ray data spanning more than a decade are now accessible. The availability of a homogeneous dataset from a large sample of blazars enables a wide range of analyses designed to characterize gamma-ray variability in these sources. Accordingly, the main objective of this study is to explore potential correlations between the variability characteristics and intrinsic physical properties of the sources, including black hole mass, luminosity, and accretion rates. The paper is structured as follows: Section 2 presents the sample and method; Section 3 presents the results and discussion; and Section 4 provides a summary and the conclusions derived from the statistical analysis. Throughout this paper, we assume a cosmology with $H_0 = 70\,\mathrm{km\,s^{-1}\,Mpc^{-1}}$, $\Omega_m = 0.3$, and $\Omega_\Lambda = 0.7$.

## 2. The Sample and Method

### 2.1. The Sample

To investigate the relationship between the amplitude of gamma-ray variability and black hole mass, luminosity, and accretion rate, we collected a sample of jetted AGNs with well-documented black hole masses and accretion-disk luminosities. First, we considered the sample of V. S. Paliya et al. (2021), in which they provided a comprehensive catalog of central engine properties, including black hole mass and accretion-disk luminosity, for a sample of 1077 blazars detected by Fermi-LAT. At the same time, V. S. Paliya et al. (2021) derived the synchrotron and inverse-Compton (IC) peak frequencies as well as the corresponding peak frequency luminosities by fitting a second-degree polynomial to the spectral energy distribution (SED). Second, we focus on sources within this dataset that exhibit detectable gamma-ray light curves. Finally, we found that 387 out of 1077 sources exhibited available gamma-ray light curves from the Fermi-LAT public Light Curve Repository[7] (S. Abdollahi et al. 2023). These 387 sources include 285 FSRQs, 65 BL Lacs, 26 blazar candidates of uncertain type (BCUs), and 11 narrow-line Seyfert 1 (NLS1) galaxies.

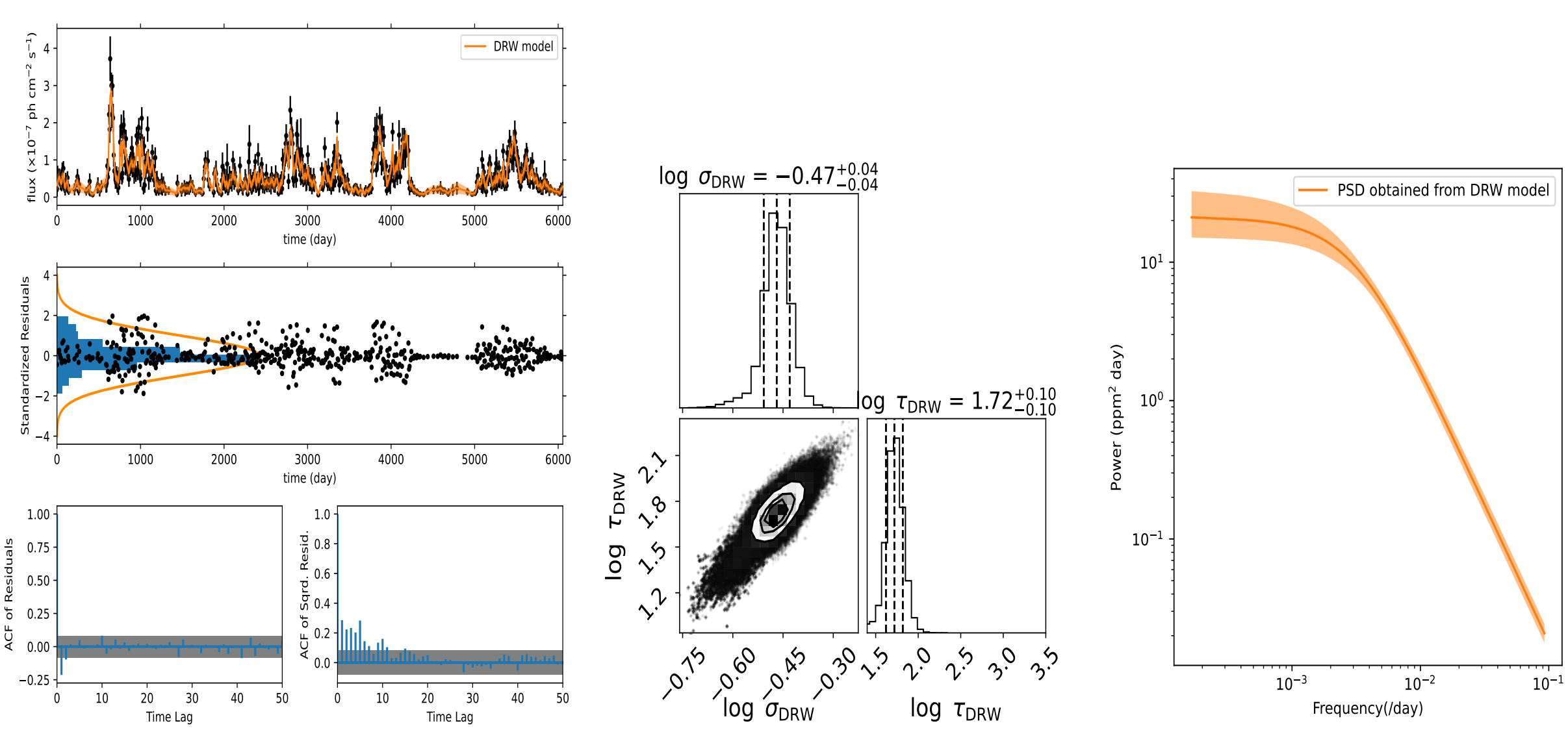

**Figure 1.** The fitting results of 4FGLJ1033.9+6050 are presented as examples. The left panel displays the LAT light curve (black points), along with the modeled light curve (orange line). We present the standardized residuals (black points), the corresponding probability density of these residuals (blue histogram), and the best-fit normal distribution (orange solid line). We also display the ACF of the residuals, along with the 95% confidence limits of white noise represented by the gray region. The middle panel is the posterior distribution of the mode parameters for the DRW. The right panel is the PSD of the DRW. The shaded region represents the $1\sigma$ confidence interval.

### 2.2. The DRW Model

The DRW process is described by a first-order stochastic differential equation (see the details in B. C. Kelly et al. 2009; J. Moreno et al. 2019). It represents a competition between a process seeking to maintain an equilibrium state and a perturbation driving the system away from stability. This relationship is often alternatively expressed in the form of a Langevin equation:

$$\left[\frac{d}{dt} + \frac{1}{\tau_{\rm DRW}}\right] y(t) = \sigma_{\rm DRW}\, \epsilon(t). \tag{1}$$

$\tau_{\rm DRW}$ is the characteristic damping timescale of the DRW process, and $\sigma_{\rm DRW}$ is representing the amplitude of the random perturbations.

We perform Gaussian process regression to fit a DRW model to the observed light curves based on the implementation provided in the `celerite` package. A Gaussian process model is fully defined by its covariance function, which is commonly known as the kernel function. The covariance function for a DRW is

$$k(t_{nm}) = 2\sigma_{\rm DRW}^2 \exp(-t_{nm}/\tau_{\rm DRW}), \tag{2}$$

where $t_{nm} = |t_n - t_m|$ denotes the time lag between the measurements $n$ and $m$. This formulation corresponds directly to the definition of the structure function:

$$\begin{aligned} {\rm SF}^2 &= {\rm SF}_\infty^2(1 - {\rm ACF}(t_{nm})) \\ &= 2\sigma_{\rm DRW}^2(1 - \exp(-t_{nm}/\tau_{\rm DRW})), \end{aligned} \tag{3}$$

where the asymptotic variability amplitude is expressed as ${\rm SF}_\infty = \sqrt{2}\,\sigma_{\rm DRW}$ (C. J. Burke et al. 2021), and the autocorrelation function (ACF) is formulated as ${\rm ACF}(t_{nm}) = \exp(-t_{nm}/\tau_{\rm DRW})$. The PSD is expressed as (D. Foreman-Mackey et al. 2017)

$$S(\omega) = \sqrt{\frac{2}{\pi}}\frac{2\sigma_{\rm DRW}^2\,\tau_{\rm DRW}}{1 + (\omega/\tau_{\rm DRW})^2}. \tag{4}$$

The DRW PSD exhibits a broken-PL form, where the spectral index transitions from 0 at low frequencies to –2 at high frequencies. The break frequency $f_{\rm b}$ corresponds to the characteristic damping timescale, defined as $\tau_{\rm DRW} = \frac{1}{2\pi f_{\rm b}}$ (H. Zhang et al. 2022).

Because in the sample of 387, some of the sources would be bright and some would be faint, it was chosen to have a common binning mostly of 3 days. We apply the DRW model to fit the gamma-ray 3 day binned light curves of 387 AGNs. An illustrative example is presented in Figure 1. However, it is crucial to take into account how the constrained duration of a light curve may impact the outcomes of the modeling. An inadequate length has the potential to skew the measurement of the damping timescale, thereby introducing possible inaccuracies. To obtain reliable measurements of the damping timescale, we adopt the following selection criteria from C. J. Burke et al. (2021) and H. Zhang et al. (2022): (1) the damping timescale $\tau_{\rm DRW} < 0.1\times$ the observational baseline;

[7] https://fermi.gsfc.nasa.gov/ssc/data/access/lat/LightCurveRepository/about.html

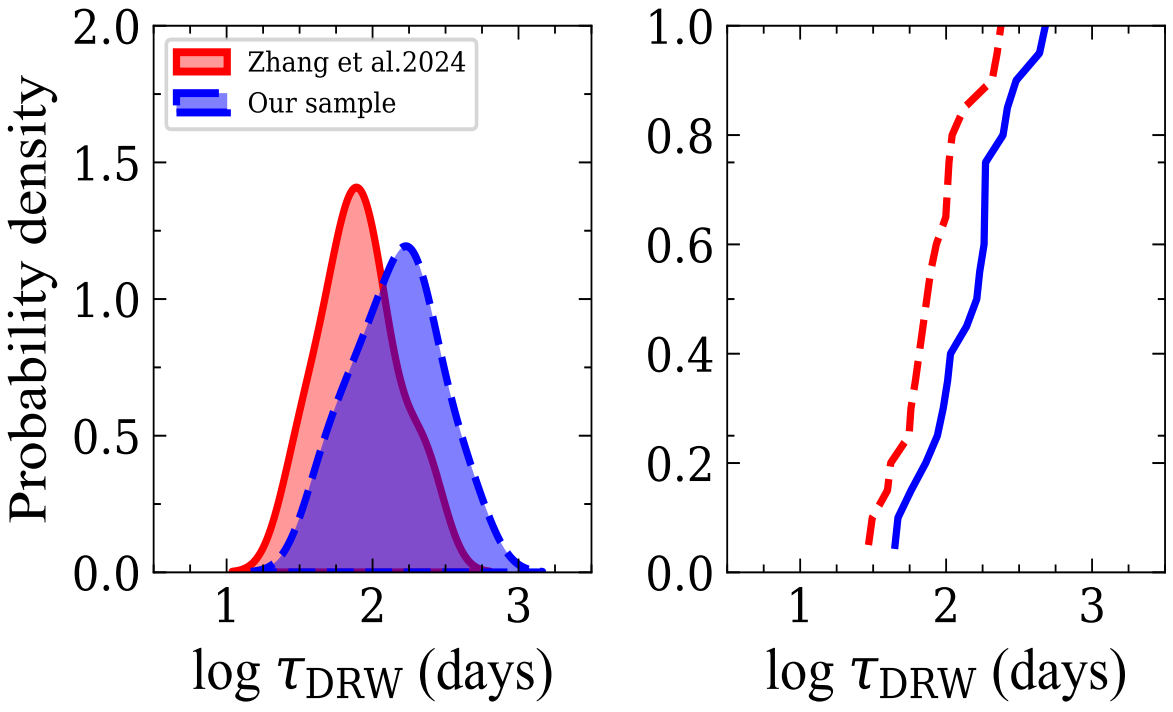

**Figure 2.** The distributions of damping timescales (left) and their cumulative distributions (right) for jetted AGNs. The red shows the damping timescales obtained by H. Zhang et al. (2024). The blue shows the damping timescales obtained by us.

and (2) the damping timescale $\tau_{\rm DRW}$ > the mean cadence. Applying the reliability criteria in the estimation of the variability timescale as mentioned above, we determined that the damping timescales and amplitude of variability for 195 out of 387 sources are reliable, consisting of 22 BL Lacs, 157 FSRQs, 7 BCUs, and 9 NLS1s. To assess the quality of the DRW model, we analyzed the probability densities of the ACF of the residuals and squared residuals for all jet AGNs. The lag values in the ACF always fall within the 95% confidence interval expected for white noise. In addition, the distribution of residuals was fitted to a normal distribution. The distribution showed good agreement with a normal distribution, characterized by the mean value of residuals being close to zero (see the left panel of Figure 1). H. Zhang et al. (2024) also used the DRW model to fit the 15 day binned light curves of 34 blazars. We find that there are 20 sources in the sample of H. Zhang et al. (2024) that are also present in our sample. We compare the damping timescale obtained by H. Zhang et al. (2024) with the damping timescale obtained by us for these 20 sources. Through the Kolmogorov–Smirnov (KS) test ($p$ = 0.03, significance level $p$ < 0.05; see Figure 2), we find a significant difference in the distribution of our damping timescale and those of H. Zhang et al. (2024). This result may indicate that the binning of the light curve has an impact on the variability. Z. Shah et al. (2025) investigated the impact of binning on the variability of gamma-ray light curves using blazar samples and found that binning affects the observed variability, with longer bins smoothing short-term fluctuations.

### *2.3. The Luminosity*

To investigate the origin of gamma-ray variability, we compiled the $B$-band optical flux, 1.4 GHz radio flux, 0.3–10 keV X-ray flux, and gamma-ray flux from 195 sources. The $B$-band optical flux data were obtained from the NASA/IPAC Extragalactic Database. The 1.4 GHz radio flux values are sourced from the NRAO VLA Sky Survey (J. J. Condon et al. 1998). The 0.3–10 keV X-ray flux measurements were extracted from the Swift X-Ray Telescope Point Source Catalog (P. A. Evans et al. 2020). The gamma-ray flux data are taken from the catalog of AGNs detected by Fermi-LAT (S. Abdollahi et al. 2022; M. Ajello et al. 2022). The fluxes across these bands were K-corrected using the relation $S_\nu = S_{{\rm obs},\nu}(1+z)^{\alpha-1}$, where $\alpha$ denotes the spectral index. For the $B$ band, the spectral index is set to $\alpha = 0.3$ (S. Rakshit & C. S. Stalin 2017; G. Gürkan et al. 2019), while for the radio band, it is $\alpha = 0.8$ (K. W. Cavagnolo et al. 2010). The $B$-band optical luminosity, 1.4 GHz radio luminosity, 0.3–10 keV X-ray luminosity, and gamma-ray luminosity were computed using the formula $L_\nu = 4\pi d_L^2 S_\nu$, where $d_L$ represents the luminosity distance. We calculated the radio loudness, X-ray loudness, and gamma-ray loudness using the following flux ratios: $R_{\rm radio} = S_{\rm radio}/S_{\rm B}$ (K. I. Kellermann et al. 1989), $R_{\rm X-ray} = S_{\rm X-ray}/S_{\rm B}$ (M. Gupta & M. Sikora 2018; W.-y. Kang et al. 2018), and $R_\gamma = S_\gamma/S_{\rm B}$ (T. G. Arshakian et al. 2012). The relevant data are listed in Tables 1 and 2. These tables are available in their entirety in machine-readable form.

## 3. Results and Discussion

### *3.1. The Distributions of Amplitudes of Variability*

A. A. Abdo et al. (2010b) analyzed 106 gamma-ray light curves using 11 months of Fermi observatory data and found that FSRQs display a greater amplitude of gamma-ray variability than other AGN types. Likewise, M. Ackermann et al. (2011) observed that FSRQs exhibit stronger flux variability than BL Lacs, based on an analysis of sources in the second LAT AGN catalog. Therefore, we compare the variability amplitudes of FSRQs and BL Lacs. The distributions of the variability amplitudes for jetted AGNs are presented in panel (a) of Figure 3. The mean variability amplitudes for FSRQs, BL Lacs, BCUs, and NLS1s are $\overline{\log {\rm SF}_{\infty,{\rm FSRQs}}} = -0.26 \pm 0.32$, $\overline{\log {\rm SF}_{\infty,{\rm BLLacs}}} = -0.57 \pm 0.41$, $\overline{\log {\rm SF}_{\infty,{\rm BCU}}} = -0.27 \pm 0.22$, and $\overline{\log {\rm SF}_{\infty,{\rm NLS1}}} = -0.35 \pm 0.28$, respectively. The average amplitude of variability for FSRQs is significantly greater than that for BL Lacs. A two-sample KS test indicates that the amplitude distributions of FSRQs and BL Lacs are statistically different at the 95% confidence level, with a test statistic of 0.502 and a $p$-value of $5.09 \times 10^{-5}$. Due to the small number of BCU and NLS1 sources (less than 10), we did not compare the differences in the distributions of physical parameters between FSRQs, BL Lacs, BCUs, and NLS1s.

According to M. Ackermann et al. (2011), the enhanced variability in FSRQs relative to BL Lacs may be attributed to the position of the high-energy peak in the broadband SED with respect to the gamma-ray band. Within this band, FSRQs emit at energies above the IC peak of their SED, indicating that the emission originates from high-energy electrons with shorter cooling timescales, leading to greater variability. In contrast, BL Lacs are observed at frequencies well below their IC peak within the gamma-ray band, implying that their emission is dominated by low-energy electrons with longer cooling timescales, resulting in reduced variability. The results of this study, derived from a large sample of blazars, are consistent with earlier findings obtained from smaller samples of blazars (A. A. Abdo et al. 2010b; M. Ackermann et al. 2011; B. Rajput et al. 2020).

### *3.2. The Distributions of the Variability Damping Timescale*

The variability damping timescale is a key parameter that can be inferred from light curves, offering important constraints for theoretical models when compared to relevant physical timescales. The distributions of the variability damping timescales for our sample are presented in panel (b) of Figure 3. The mean variability damping timescales for FSRQs, BL Lacs, BCUs, and NLS1s are $\overline{\log \tau_{\rm DRW,FSRQs}} = 1.70 \pm 0.42$, $\overline{\log \tau_{\rm DRW,BLLacs}} = 2.07 \pm 0.37$, $\overline{\log \tau_{\rm DRW,BCU}} = 2.08 \pm 0.24$, and $\overline{\log \tau_{\rm DRW,NLS1}} = 1.95 \pm 0.54$, respectively. The variability damping timescale of FSRQs is shorter than that of BL Lacs. The KS test indicates that

**Table 1**
The Physical Parameters for Gamma-ray-emitting Jetted AGNs

| 4FGL Name (1) | R.A. (2) | Decl. (3) | Type (4) | $S_B$ (5) | $S_{B,error}$ (6) | $S_{0.3-10\ KeV}$ (7) | $S_{0.3-10\ KeV,error}$ (8) | $\Gamma_X$ (9) | $S_{1.4\ GHz}$ (10) | $S_{1.4\ GHz,error}$ (11) | $S_\gamma$ (12) | $S_{\gamma,error}$ (13) | $\Gamma_\gamma$ (14) | $\log L_{sy}$ (15) | $\log L_{IC}$ (16) |
|---|---|---|---|---|---|---|---|---|---|---|---|---|---|---|---|
| 4FGL J0001.5+2113 | 359.69202 | 19.92231 | NLS1 | 3.06E-04 | 1.03E-04 | 5.52E-13 | 4.6E-14 | 1.63 | 672.5 | 20.2 | 2.61E-11 | 7.9E-13 | 2.66 | 43.78 | 45.27 |
| 4FGL J0011.4+0057 | 2.87668 | 0.96441 | FSRQs | 4.42E-05 | 1.84E-05 | 1.71E-13 | 4.3E-14 | 2.00 | 167 | 5 | 5.34E-12 | 5.08E-13 | 2.33 | 45.04 | 45.60 |
| 4FGL J0028.4+2001 | 7.12424 | 20.00743 | FSRQs | 4.42E-05 | 9.80E-06 | 7.8E-12 | 5.39E-12 | 2.00 | 286.5 | 8.6 | 8.1E-12 | 5.34E-13 | 2.41 | 45.05 | 45.46 |
| 4FGL J0030.6-0212 | 7.6326 | −2.19893 | FSRQs | 4.30E-05 | 1.79E-05 | 7.8E-12 | 5.39E-12 | 2.00 | 150.6 | 4.5 | 1.72E-11 | 7.73E-13 | 2.40 | 44.96 | 46.01 |
| 4FGL J0038.2-2459 | 9.5614 | −24.98395 | FSRQs | 3.58E-05 | 1.45E-05 | 4.79E-12 | 1.385E-13 | 1.46 | 412.5 | 12.4 | 1.42E-11 | 7.09E-13 | 2.26 | 44.15 | 44.42 |
| 4FGL J0050.0-5736 | 12.4978 | −57.64093 | FSRQs | 7.20E-05 | 0.00E+00 | 1.727E-12 | 1.705E-13 | 1.60 | 2110 | 370 | 6.59E-12 | 4.62E-13 | 2.62 | 45.49 | 45.91 |
| 4FGL J0102.8+5824 | 15.69068 | 58.40309 | BCU | 3.71E-05 | 1.55E-05 | 3.173E-12 | 8.985E-14 | 1.55 | 848.2 | 25.5 | 4.34E-11 | 1.42E-12 | 2.29 | 45.02 | 45.19 |
| 4FGL J0108.6+0134 | 17.16155 | 1.58342 | FSRQs | 1.25E-04 | 5.20E-05 | 2.187E-12 | 4.617E-14 | 1.46 | 2620.7 | 78.6 | 1.27E-10 | 1.6E-12 | 2.36 | 46.04 | 46.87 |
| 4FGL J0113.4+4948 | 18.36253 | 49.80668 | FSRQs | 1.06E-04 | 2.80E-05 | 2.93E-12 | 2.077E-13 | 1.50 | 666.8 | 20 | 1.61E-11 | 9.06E-13 | 2.22 | 44.10 | 44.29 |
| 4FGL J0116.0-1136 | 19.05217 | −11.60429 | FSRQs | 1.58E-04 | 6.40E-05 | 3.597E-12 | 1.537E-13 | 1.67 | 1785 | 53.6 | 1.08E-11 | 5.39E-13 | 2.36 | 44.84 | 45.00 |
| 4FGL J0118.9-2141 | 19.73859 | −21.69171 | FSRQs | 5.77E-05 | 2.29E-05 | 7.942E-13 | 7.138E-14 | 1.64 | 447.4 | 13.4 | 3.01E-11 | 1.13E-12 | 2.26 | 45.21 | 45.84 |
| 4FGL J0132.7-1654 | 23.1812 | −16.91348 | FSRQs | 1.02E-04 | 1.50E-05 | 2.083E-12 | 9.774E-14 | 1.47 | 830.3 | 24.9 | 2.42E-11 | 7.75E-13 | 2.34 | 45.43 | 45.69 |
| 4FGL J0133.1-5201 | 23.27401 | −52.0011 | FSRQs | 1.71E-04 | 2.60E-05 | 2.54E-12 | 1.441E-13 | 1.38 | 352 | 70 | 1.84E-11 | 6.84E-13 | 2.32 | 44.28 | 45.33 |
| 4FGL J0137.0+4751 | 24.24414 | 47.85808 | FSRQs | 5.12E-05 | 2.13E-05 | 3.122E-12 | 6.587E-14 | 1.45 | 1137.5 | 34.1 | 3.64E-11 | 1.04E-12 | 2.30 | 45.47 | 45.68 |
| 4FGL J0137.6-2430 | 24.40978 | −24.51497 | FSRQs | 6.69E-04 | 8.70E-05 | 3.316E-12 | 1.165E-13 | 1.67 | 1181.2 | 40.7 | 1.36E-11 | 5.34E-13 | 2.54 | 45.13 | 45.52 |
| 4FGL J0203.7+3042 | 30.93898 | 30.69142 | FSRQs | 7.45E-04 | 2.17E-04 | 7.8E-12 | 5.39E-12 | 2.00 | 810.9 | 24.3 | 2.09E-11 | 9.44E-13 | 2.24 | 44.68 | 45.05 |
| 4FGL J0204.8+1513 | 31.21006 | 15.2364 | FSRQs | 1.07E-04 | 0.00E+00 | 8.945E-13 | 5.12E-14 | 1.31 | 4067.7 | 122 | 1.01E-11 | 5.72E-13 | 2.31 | 44.04 | 44.39 |
| 4FGL J0205.0-1700 | 31.24031 | −17.02218 | FSRQs | 2.52E-04 | 9.70E-05 | 4.994E-12 | 2.437E-13 | 1.41 | 1219.5 | 36.6 | 1.45E-11 | 6.17E-13 | 2.63 | 45.33 | 46.27 |
| 4FGL J0205.2+3212 | 31.27052 | 32.20836 | FSRQs | 7.47E-04 | 2.96E-04 | 1.246E-12 | 7.31E-14 | 1.53 | 657 | 19.7 | 1.28E-11 | 6.56E-13 | 2.66 | 45.29 | 46.06 |
| 4FGL J0210.7-5101 | 32.6925 | −51.01719 | FSRQs | 4.76E-04 | 4.60E-05 | 3.352E-12 | 3.109E-14 | 1.61 | 3490 | 150 | 5.87E-11 | 1.05E-12 | 2.34 | 45.53 | 45.92 |
| 4FGL J0217.8+0144 | 34.45398 | 1.74714 | FSRQs | 4.43E-05 | 6.70E-06 | 3.85E-12 | 1.582E-12 | 2.75 | 750.2 | 22.5 | 2.88E-11 | 9.58E-13 | 2.24 | 46.09 | 46.19 |
| 4FGL J0222.0-1616 | 35.50302 | −16.2546 | FSRQs | 3.51E-05 | 1.43E-05 | 1.056E-12 | 1.325E-13 | 1.42 | 589.8 | 17.7 | 4.37E-12 | 5.55E-13 | 2.40 | 44.27 | 44.92 |
| 4FGL J0223.2-1653 | 35.93235 | −16.94381 | FSRQs | 6.69E-05 | 2.65E-05 | 7.8E-12 | 5.39E-12 | 2.00 | 240.7 | 7.2 | 2.72E-12 | 6.59E-13 | 2.72 | 44.54 | 44.98 |
| 4FGL J0237.8+2848 | 39.46836 | 28.8025 | FSRQs | 6.10E-05 | 2.24E-05 | 3.254E-12 | 8.324E-14 | 1.56 | 2196.9 | 65.9 | 1.15E-10 | 1.49E-12 | 2.32 | 45.68 | 46.22 |
| 4FGL J0245.9-4650 | 41.50049 | −46.85479 | FSRQs | 1.43E-04 | 0.00E+00 | 2.466E-12 | 1.491E-13 | 1.58 | 1500 | 34 | 3.11E-11 | 8.89E-13 | 2.42 | 45.50 | 46.31 |
| 4FGL J0252.8-2219 | 43.19981 | −22.32374 | FSRQs | 1.50E-04 | 0.00E+00 | 8.582E-13 | 6.151E-14 | 1.38 | 443.1 | 13.3 | 4.55E-11 | 8.93E-13 | 2.36 | 45.02 | 46.27 |
| 4FGL J0253.2-5441 | 43.37158 | −54.69762 | FSRQs | 4.14E-04 | 7.50E-05 | 3.083E-12 | 1.915E-13 | 1.66 | 896 | 13 | 7.01E-12 | 4.08E-13 | 2.53 | 44.73 | 44.46 |
| 4FGL J0259.4+0746 | 44.86282 | 7.79435 | FSRQs | 1.32E-04 | 5.10E-05 | 6.031E-13 | 1.175E-13 | 2.00 | 834.3 | 25 | 1.97E-11 | 1.04E-12 | 2.20 | 44.55 | 45.15 |
| 4FGL J0301.6-7155 | 45.41019 | −71.94289 | FSRQs | 7.45E-04 | 2.17E-04 | 6.732E-13 | 8.775E-14 | 1.19 | 790 | 0 | 4.46E-12 | 4.02E-13 | 2.54 | 44.36 | 44.99 |

**Notes.** Column (1): source name of 4FGL. Column (2): the R.A. in decimal degrees. Column (3): the decl. in decimal degrees. Column (4): the types of sources. Column (5): the optical *B*-band flux, in units of Jy. Column (6): the optical *B*-band flux error, in units of Jy. Column (7): the X-ray flux in 0.3–10 keV, in units of erg $s^{-1}$ $cm^{-2}$. Column (8): the X-ray flux error in 0.3–10 keV, in units of Jy. Column (9): the photon spectral index. Column (10): the 1.4 GHz flux, in units of mJy. Column (11): the 1.4 GHz flux error, in units of mJy. Column (12): the gamma-ray flux, in units of erg $s^{-1}$ $cm^{-2}$. Column (13): the gamma-ray flux error, in units of erg $s^{-1}$ $cm^{-2}$. Column (14): photon spectral index for gamma rays. Columns (15)–(16): the synchronous peak frequency and IC peak frequency luminosity, measured in units of erg $s^{-1}$.

(This table is available in its entirety in machine-readable form in the online article.)

**Table 2**
The DRW Model Parameter, Black Hole Mass, and Accretion-disk Luminosity of Jetted AGNs

| Name (1) | $z$ (2) | $\log \sigma_{\rm DRW}$ (3) | $\log \tau_{\rm DRW}$ (4) | $\log M_{\rm BH}$ (5) | $\log L_{\rm disk}$ (6) | $\log \delta$ (7) |
|---|---|---|---|---|---|---|
| 4FGL J0001.5+2113 | 0.439 | $-0.18^{+0.04}_{-0.04}$ | $1.28^{+0.17}_{-0.17}$ | 7.54 ± 0.07 | 44.65 ± 0.02 | 0.57 |
| 4FGL J0011.4+0057 | 1.491 | $-0.5^{+0.06}_{-0.05}$ | $1.9^{+0.34}_{-0.24}$ | 8.66 ± 0.05 | 45.71 ± 0.02 | 0.66 |
| 4FGL J0028.4+2001 | 1.553 | $0.05^{+0.06}_{-0.05}$ | $2.18^{+0.14}_{-0.12}$ | 8.43 ± 0.08 | 45.57 ± 0.04 | 0.66 |
| 4FGL J0030.6-0212 | 1.804 | $-0.51^{+0.04}_{-0.04}$ | $1.69^{+0.18}_{-0.16}$ | 8.66 ± 0.28 | 45.98 ± 0.02 | 0.69 |
| 4FGL J0038.2-2459 | 0.498 | $-0.09^{+0.02}_{-0.02}$ | $1.18^{+0.08}_{-0.07}$ | 8.14 ± 0.23 | 44.97 ± 0.07 | 0.62 |
| 4FGL J0050.0-5736 | 1.797 | $-0.57^{+0.1}_{-0.07}$ | $2.47^{+0.32}_{-0.27}$ | 9.06 ± 0.34 | 46.88 ± 0.02 | 0.63 |
| 4FGL J0102.8+5824 | 0.644 | $-0.87^{+0.1}_{-0.07}$ | $2.47^{+0.27}_{-0.2}$ | 9.01 ± 0.25 | 46.04 ± 0.15 | 0.67 |
| 4FGL J0108.6+0134 | 2.109 | $0.42^{+0.06}_{-0.05}$ | $2.25^{+0.13}_{-0.1}$ | 9.64 ± 0.11 | 46.62 ± 0.03 | 0.77 |
| 4FGL J0113.4+4948 | 0.389 | $-0.9^{+0.08}_{-0.07}$ | $2.32^{+0.22}_{-0.17}$ | 8.23 ± 0.1 | 44.78 ± 0.03 | 0.62 |
| 4FGL J0116.0-1136 | 0.67 | $-0.61^{+0.05}_{-0.05}$ | $1.64^{+0.24}_{-0.18}$ | 8.65 ± 0.16 | 45.69 ± 0.03 | 0.62 |
| 4FGL J0118.9-2141 | 1.165 | $-0.71^{+0.05}_{-0.04}$ | $1.89^{+0.16}_{-0.14}$ | 8.88 ± 0.18 | 45.71 ± 0.03 | 0.70 |
| 4FGL J0132.7-1654 | 1.02 | $-0.63^{+0.05}_{-0.05}$ | $2.08^{+0.28}_{-0.25}$ | 8.66 ± 0.66 | 46.6 ± 0.05 | 0.67 |
| 4FGL J0133.1-5201 | 0.925 | $-0.15^{+0.02}_{-0.02}$ | $1.19^{+0.07}_{-0.06}$ | 8.5 ± 0.4 | 46.32 ± 0.05 | 0.66 |
| 4FGL J0137.0+4751 | 0.859 | $-0.43^{+0.04}_{-0.03}$ | $1.74^{+0.1}_{-0.09}$ | 8.33 ± 0.29 | 45.6 ± 0.05 | 0.68 |
| 4FGL J0137.6-2430 | 0.835 | $-0.73^{+0.09}_{-0.07}$ | $2.31^{+0.39}_{-0.32}$ | 8.94 ± 0.22 | 46.42 ± 0.08 | 0.61 |
| 4FGL J0203.7+3042 | 0.761 | $-0.47^{+0.15}_{-0.09}$ | $2.67^{+0.36}_{-0.22}$ | 8.41 ± 0.1 | 45.01 ± 0.07 | 0.67 |
| 4FGL J0204.8+1513 | 0.407 | $-0.53^{+0.16}_{-0.1}$ | $2.76^{+0.39}_{-0.24}$ | 7.75 ± 0.36 | 43.83 ± 0.16 | 0.59 |
| 4FGL J0205.0-1700 | 1.737 | $-0.21^{+0.03}_{-0.03}$ | $1.58^{+0.1}_{-0.09}$ | 9.5 ± 0.14 | 47.23 ± 0.04 | 0.65 |
| 4FGL J0205.2+3212 | 1.466 | $-0.64^{+0.07}_{-0.06}$ | $2.18^{+0.27}_{-0.23}$ | 8.83 ± 0.22 | 46.55 ± 0.1 | 0.63 |
| 4FGL J0210.7-5101 | 1.003 | $0.07^{+0.07}_{-0.05}$ | $2.29^{+0.14}_{-0.11}$ | 9.12 ± 0.2 | 46.05 ± 0.19 | 0.70 |
| 4FGL J0217.8+0144 | 1.715 | $-0.58^{+0.09}_{-0.07}$ | $2.48^{+0.23}_{-0.16}$ | 9.22 ± 0.21 | 45.85 ± 0.04 | 0.74 |
| 4FGL J0222.0-1616 | 0.698 | $-0.98^{+0.17}_{-0.11}$ | $2.74^{+0.53}_{-0.37}$ | 7.58 ± 0.48 | 45.34 ± 0.05 | 0.59 |
| 4FGL J0223.2-1653 | 1.015 | $-0.54^{+0.05}_{-0.05}$ | $1.32^{+0.18}_{-0.16}$ | 8.13 ± 0.24 | 45.04 ± 0.07 | 0.56 |
| 4FGL J0237.8+2848 | 1.206 | $0.13^{+0.04}_{-0.04}$ | $1.93^{+0.1}_{-0.08}$ | 9.13 ± 0.25 | 46.4 ± 0.07 | 0.73 |
| 4FGL J0245.9-4650 | 1.385 | $-0.4^{+0.06}_{-0.05}$ | $2.13^{+0.15}_{-0.13}$ | 8.78 ± 0.19 | 46.58 ± 0.07 | 0.69 |
| 4FGL J0252.8-2219 | 1.455 | $-0.31^{+0.04}_{-0.04}$ | $1.92^{+0.11}_{-0.09}$ | 8.98 ± 0.72 | 45.39 ± 0.15 | 0.71 |
| 4FGL J0253.2-5441 | 0.539 | $-0.63^{+0.12}_{-0.08}$ | $2.46^{+0.58}_{-0.59}$ | 8.75 ± 0.1 | 46.02 ± 0.02 | 0.56 |
| 4FGL J0259.4+0746 | 0.893 | $-0.72^{+0.07}_{-0.06}$ | $2.2^{+0.25}_{-0.21}$ | 7.49 ± 0.12 | 44.74 ± 0.06 | 0.68 |
| 4FGL J0301.6-7155 | 0.823 | $-0.09^{+0.07}_{-0.06}$ | $2.22^{+0.25}_{-0.22}$ | 8.78 ± 0.14 | 45.57 ± 0.02 | 0.58 |
| 4FGL J0309.0+1029 | 0.862 | $-0.52^{+0.06}_{-0.05}$ | $1.68^{+0.2}_{-0.18}$ | 7.77 ± 0.16 | 44.76 ± 0.05 | 0.65 |
| 4FGL J0309.9-6058 | 1.482 | $-0.45^{+0.04}_{-0.04}$ | $1.84^{+0.11}_{-0.09}$ | 8.87 ± 0.34 | 45.85 ± 0.09 | 0.67 |
| 4FGL J0325.7+2225 | 2.066 | $-0.61^{+0.04}_{-0.04}$ | $1.57^{+0.09}_{-0.06}$ | 9.04 ± 0.16 | 46.86 ± 0.09 | 0.67 |
| 4FGL J0339.5-0146 | 0.852 | $-0.23^{+0.03}_{-0.03}$ | $1.52^{+0.09}_{-0.08}$ | 8.7 ± 0.32 | 46.03 ± 0.07 | 0.69 |
| 4FGL J0348.5-2749 | 0.991 | $0.3^{+0.03}_{-0.03}$ | $1.69^{+0.07}_{-0.06}$ | 8.54 ± 0.1 | 45.95 ± 0.05 | 0.71 |
| 4FGL J0401.7+2112 | 0.837 | $-0.58^{+0.07}_{-0.07}$ | $1.71^{+0.6}_{-0.33}$ | 8.43 ± 0.36 | 45.15 ± 0.28 | 0.62 |
| 4FGL J0403.9-3605 | 1.428 | $0.03^{+0.04}_{-0.03}$ | $1.83^{+0.1}_{-0.09}$ | 9.65 ± 0.08 | 46.62 ± 0.09 | 0.69 |
| 4FGL J0405.6-1308 | 0.571 | $-0.61^{+0.17}_{-0.08}$ | $2.54^{+0.88}_{-0.82}$ | 8.89 ± 0.21 | 46.58 ± 0.05 | 0.57 |
| 4FGL J0423.3-0120 | 0.916 | $-0.51^{+0.06}_{-0.05}$ | $2.15^{+0.17}_{-0.14}$ | 8.4 ± 0.23 | 45.65 ± 0.05 | 0.68 |
| 4FGL J0428.6-3756 | 1.11 | $-0.05^{+0.05}_{-0.04}$ | $2.03^{+0.1}_{-0.08}$ | 8.77 ± 0.37 | 44.9 ± 0.26 | 0.79 |
| 4FGL J0433.6-6030 | 0.93 | $-1.01^{+0.12}_{-0.1}$ | $2.54^{+0.54}_{-0.42}$ | 8.4 ± 0.32 | 45.85 ± 0.04 | 0.63 |
| 4FGL J0433.6+2905 | 0.91 | $-0.71^{+0.11}_{-0.07}$ | $2.56^{+0.29}_{-0.21}$ | 8.25 ± 0.55 | 44.93 ± 0.13 | 0.72 |
| 4FGL J0442.6-0017 | 0.845 | $-0.16^{+0.05}_{-0.04}$ | $2.09^{+0.12}_{-0.1}$ | 8.29 ± 0.17 | 46.14 ± 0.03 | 0.66 |
| 4FGL J0449.1+1121 | 2.153 | $-0.16^{+0.05}_{-0.05}$ | $2.1^{+0.15}_{-0.11}$ | 8.63 ± 0.41 | 46.08 ± 0.13 | 0.70 |
| 4FGL J0457.0-2324 | 1.003 | $0.02^{+0.03}_{-0.03}$ | $1.72^{+0.07}_{-0.06}$ | 8.33 ± 0.14 | 45.6 ± 0.06 | 0.75 |
| 4FGL J0501.2-0158 | 2.286 | $-0.17^{+0.03}_{-0.03}$ | $1.63^{+0.08}_{-0.07}$ | 8.66 ± 0 | 46.32 ± 0 | 0.75 |
| 4FGL J0505.3+0459 | 0.96 | $-0.24^{+0.05}_{-0.04}$ | $2.03^{+0.13}_{-0.11}$ | 8.54 ± 0.26 | 45.92 ± 0.04 | 0.69 |

**Notes.** Column (1): name. Column (2): the redshift. Column (3): the amplitude of variability, in units of magnitude. Column (4): the damping timescale, in units of days. Column (5): the black hole mass. Column (6): the accretion-disk luminosity, in units of erg $s^{-1}$. Column (7): the Doppler factor.

(This table is available in its entirety in machine-readable form in the online article.)

the variability damping timescale distributions of FSRQs and BL Lacs are statistically different at the 95% confidence level, with a test statistic of 0.44 and a $p$-value of $6.0 \times 10^{-4}$. Although blazars have traditionally been classified into FSRQs and BL Lacs based on the presence and width of the emission lines in their optical spectra, G. Ghisellini et al. (2009) argue that a more fundamental distinction between these subclasses should be based on their intrinsic physical properties. The PSDs associated with the external Compton (EC) process—responsible for gamma-ray emission in FSRQs—and the synchrotron self-Compton (SSC) mechanism—responsible for gamma-ray radiation in BL Lacs—exhibit different break frequencies (J. L. Ryan et al. 2019).

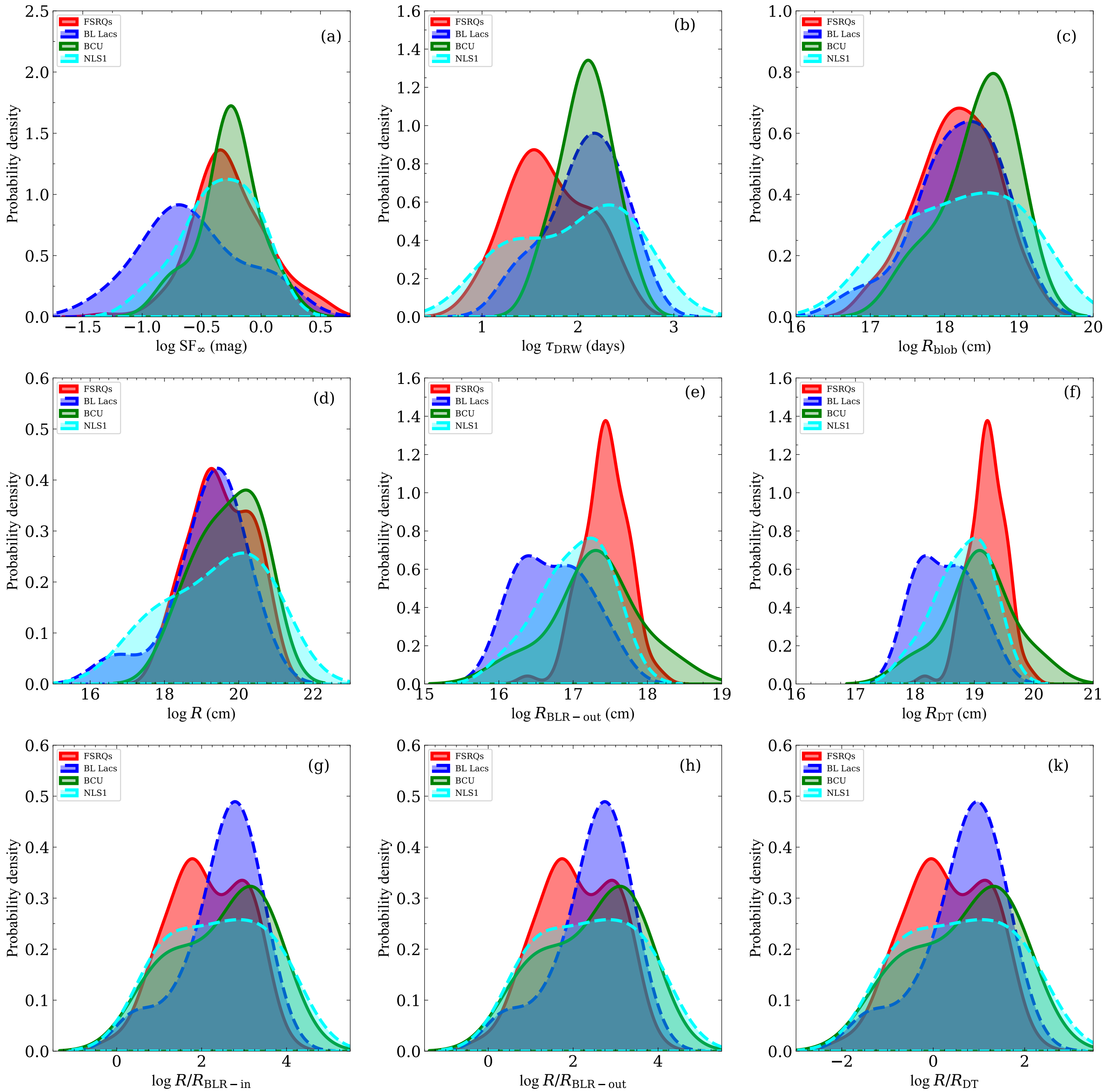

**Figure 3.** (a) The distributions of amplitudes of variability; (b) the distributions of variability timescales in the rest frame; (c) the size of the emitting region in jets ($R_{\rm blob}$); (d) the distance of the emission region from the central SMBH ($R$); (e) the outer radius of the BLR ($R_{\rm BLR-out}$); (f) the radius of the dusty torus ($R_{\rm DT}$) for jetted AGNs; (g) the ratio of $R/R_{\rm BLR-in}$; (h) $R/R_{\rm BLR-out}$; and (k) $R/R_{\rm DT}$. The red histogram represents FSRQs. The blue histogram represents BL Lacs. The green histogram represents BCUs. The cyan histogram represents NLS1s.

Within this theoretical framework, distinct variability damping timescales in the gamma-ray band are expected.

Blazar variability, ranging from minutes to years, is driven by particle acceleration mechanisms in relativistic jets, with faster (minute-scale) flares pointing to magnetic reconnection and slower (day-to-month) variability often attributed to diffusive shock acceleration (e.g., A. Shukla & K. Mannheim 2020; R. Khatoon et al. 2024; C.-R. Hu et al. 2025; C. M. Raiteri 2025; C. K. Das et al. 2026). The variability damping timescales of our sources are of the order of 100 days —see Table 2. This damping timescale may imply that the diffusive shock acceleration plays an important role in the variability of gamma-ray emission. At the same time, the variability damping timescale of approximately 100 days may favor associating the observed variability with intrinsic processes in the accretion disk, which can be naturally interpreted within the framework of inner propagating fluctuation models (Y. E. Lyubarskii 1997). Such a model inherently produces a log-normal flux distribution, analogous to the "exponential model" of P. Uttley et al. (2005). Since the jet-launching region lies near the accretion disk, variability in the disk is expected to be imprinted on jet emission (J. Kataoka et al. 2001). Many authors have studied the variability damping timescales of optical observations from the accretion disks of AGNs (S. Collier & B. M. Peterson 2001; B. C. Kelly et al. 2009; C. L. MacLeod et al. 2010; T. Simm et al. 2016; C. J. Burke et al. 2021; K. L. Suberlak et al. 2021; H. Zhang et al. 2023). Recently, analogous research has been carried out in the submillimeter wavelength region (Y. Chen et al. 2023), X-rays (H. Zhang et al. 2024), and gamma rays (e.g., J. L. Ryan et al. 2019; H. Zhang et al. 2022, 2024; A. Sharma et al. 2026).

Correlations between the variability damping timescales and black hole mass can allow us to better constrain the physical origin of the variability damping timescales. A linear correlation of the variability damping timescales and black hole mass has been identified in the X-ray band among diverse black hole accretion systems, from Galactic black hole X-ray binaries to blazars (e.g., I. McHardy 1988; I. M. McHardy et al. 2004; R. Chatterjee et al. 2018). This trend has also been detected in optical observations (e.g., B. C. Kelly et al. 2009; C. L. MacLeod et al. 2010; C. J. Burke et al. 2021; D. Xiong et al. 2025). Recently, some authors have also investigated the correlation between variability damping timescales and black hole masses in the gamma-ray band (e.g., J. L. Ryan et al. 2019; H. Zhang et al. 2022, 2024; A. Sharma et al. 2026). We also study the relationship between black hole mass and the intrinsic variability damping timescale and compare it with the work of H. Zhang et al. (2024). The intrinsic variability damping timescale can be calculated by taking into account the Doppler beaming effect and the redshift of each source as $\tau^{\rm in}_{\rm DRW} = \frac{\tau_{\rm DRW}\delta}{1+z}$, where $\delta$ signifies the Doppler factor, and $z$ indicates the source's redshift. The Doppler factors used in this study were obtained by cross-referencing with the dataset from I. Liodakis et al. (2018), which provides variability-derived estimates of Doppler factors in the radio band for the largest sample of radio-loud blazars. A total of 129 jetted AGNs in our sample have associated Doppler factor measurements. For the remaining 66 jetted AGNs, we compute their Doppler factors based on gamma-ray luminosities. The beaming factor is derived from the gamma-ray luminosity using the method outlined in R. S. Nemmen et al. (2012) and Y. Chen et al. (2023). Specifically, the relationship between the beaming factor ($f_{\rm b}$) and gamma-ray luminosity ($L_\gamma$) is given by Y. Chen et al. (2023) as: $\log f_{\rm b} = (-0.21 \pm 0.03)\log {\rm L}_\gamma + (7.67 \pm 1.54)$. The Doppler factor can then be estimated using the expression $\log\delta = \log f_{\rm b}/(-2 - \Gamma)$, where $\Gamma$ denotes the photon index.

The relationship between black hole mass and the intrinsic variability damping timescale for jetted and nonjetted accretion systems is shown in Figure 4. We conduct a linear regression analysis using LINMIX (B. C. Kelly 2007). We find a significant correlation between black hole mass and the intrinsic variability damping timescale for jetted and nonjetted accretion systems, respectively. The best-fit linear equation for nonjetted accretion systems is ($r = 0.98$, $p = 1.0 \times 10^{-49}$)

$$\log\tau^{\rm in}_{\rm DRW} = (0.49 \pm 0.01)\log M_{\rm BH} + (-1.88 \pm 0.07). \quad (5)$$

The best-fit linear equation for jetted accretion systems is ($r = 0.78$, $p = 4.5 \times 10^{-43}$)

$$\log\tau^{\rm in}_{\rm DRW} = (0.54 \pm 0.03)\log M_{\rm BH} + (-2.00 \pm 0.27). \quad (6)$$

H. Zhang et al. (2024) found that the slope of the relationship between black hole mass and the intrinsic variability damping timescale for jetted accretion systems is $0.57 \pm 0.02$. Our results are consistent with those of H. Zhang et al. (2024).

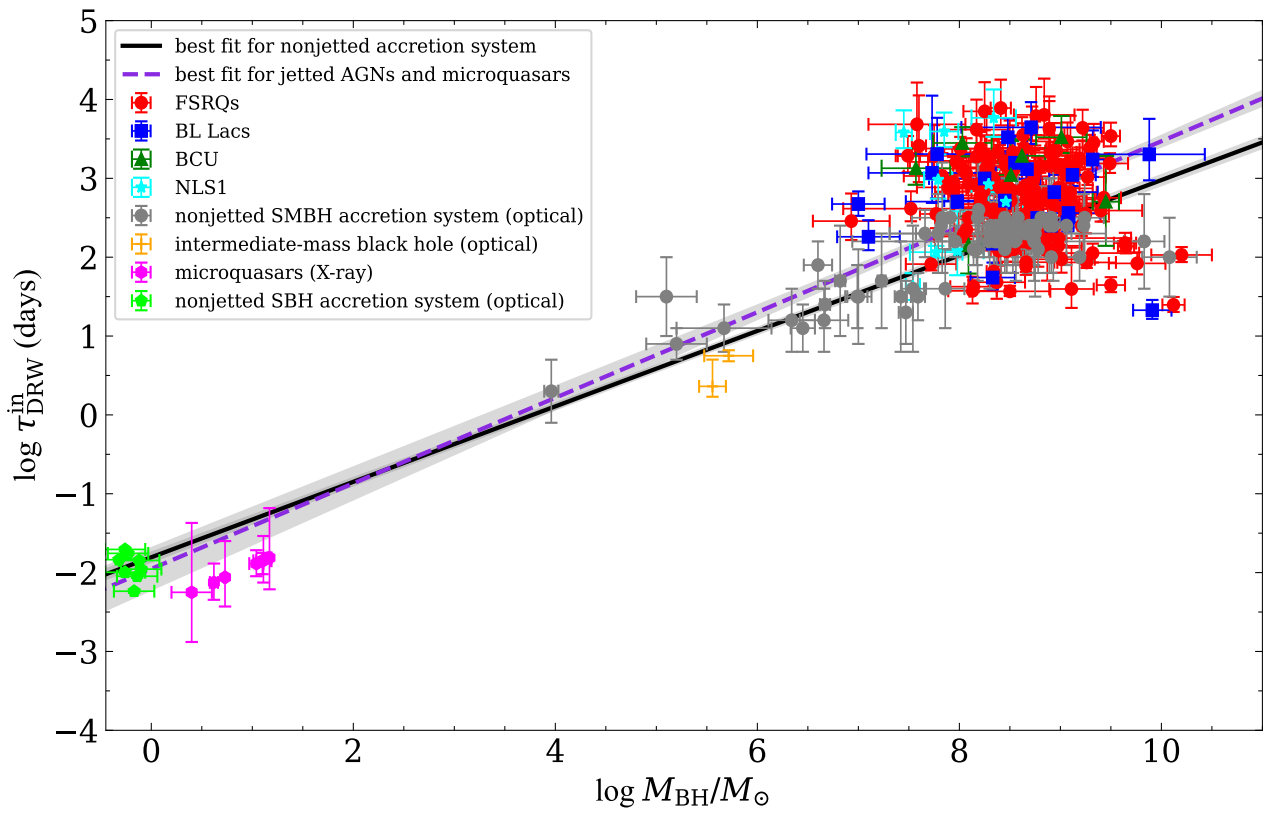

**Figure 4.** Relation between black hole mass and the intrinsic variability damping timescale for jetted and nonjetted AGNs after correcting for the beaming effect and redshift. The gray and green data points for nonjetted systems are from C. J. Burke et al. (2021). The magenta data points come from the work of H. Zhang et al. (2024). The orange data points come from the work of Z.-B. Su et al. (2024). SBH: stellar-mass black hole.

Meanwhile, we found that the jetted systems and the nonjetted systems have similar slopes (∼0.50). The observed characteristics may arise from physical processes in the accretion disk that drive the variability of the jet. At the same time, these findings shed new light on the origins of jets and the physical processes governing their emission. Moreover, these investigations confirm previous findings—namely, that the characteristics and generation mechanisms of relativistic jets could be independent of the black hole mass (A. Sharma et al. 2026).

Determining the variability damping timescale from blazar light curves is of critical importance, as it provides essential constraints on the size of the emission region within blazar jets. Based on the variability damping timescales derived from blazar gamma-ray emission data, the size of the emitting region—assumed to be a spherical plasma blob—can be estimated using the following equation (A. Sharma et al. 2026):

$$R_{\rm blob} \leqslant \frac{ct_{\rm var}\delta}{1+z}\ {\rm cm}, \quad (7)$$

where $R_{\rm blob}$ denotes the radius of the emission region, $c$ represents the speed of light, and $t_{\rm var}$ refers to the observed variability timescale.

The distribution of the size of the emitting region in jets is presented in panel (c) of Figure 3. The mean sizes of the emitting regions in the jets of FSRQs, BL Lacs, BCUs, and NLS1s are $\overline{\log R_{\rm blob,FSRQs}} = 18.17 \pm 0.53$, $\overline{\log R_{\rm blob,BLLacs}} = 18.18 \pm 0.55$, $\overline{\log R_{\rm blob,BCU}} = 18.45 \pm 0.44$, and $\overline{\log R_{\rm blob,NLS1}} = 18.22 \pm 0.72$, respectively. The KS test indicates that there is no statistically significant difference in the distribution of emitting region sizes between FSRQs and BL Lacs (statistic = 0.104, $p$ = 0.968). Furthermore, the distance of the emission region from the central supermassive black hole (SMBH), denoted as $R$, can be estimated using the following expression:

$$R \leqslant \frac{2ct_{\rm var}\delta^2}{1+z}\ {\rm cm}. \quad (8)$$

The distribution of the distance of the emission region from the central SMBH is presented in panel (d) of Figure 3. The mean distances of the emission region from the central SMBH for FSRQs, BL Lacs, BCUs, and NLS1s are $\overline{\log R_{\rm FSRQs}} = 19.52 \pm 0.82$, $\overline{\log R_{\rm BLLacs}} = 19.16 \pm 0.99$, $\overline{\log R_{\rm BCU}} = 19.71 \pm 0.78$, and $\overline{\log R_{\rm NLS1}} = 19.39 \pm 1.24$, respectively. The KS test indicates that there is no statistically significant difference in the distribution of distances of the emission region from the central SMBH between FSRQs and BL Lacs (statistic = 0.206, $p = 0.33$).

The dominant emission mechanism and the precise spatial localization of gamma-ray emission regions in AGNs have long remained subjects of extensive scientific debate. The gamma-ray emission detected from AGNs is primarily attributed to IC scattering—a process in which relativistic electrons within the jet upscatter low-energy (soft) photons to higher energies. The low-energy seed photons involved in this emission mechanism may originate from synchrotron radiation produced within the jet itself—referred to as SSC—or from external photons emitted by the accretion disk, broad-line region (BLR), or dusty torus, a mechanism known as EC. Previous studies have proposed that the gamma-ray-emitting region may be located either within the BLR at subparsec scales or at greater distances exceeding 1 parsec, beyond the BLR (S. Gulati et al. 2024). Identifying the exact location of gamma-ray emissions is crucial for elucidating the underlying emission mechanism responsible for the observed high-energy radiation. We therefore estimate the radii of both the BLR and the dusty torus. The inner radius of the BLR is determined based on the following relationship (A. Sharma et al. 2026):

$$R_{\rm BLR-in} = 3 \times 10^{17} \times \sqrt{\frac{L_{\rm disk}}{10^{46}}}\ {\rm cm}, \tag{9}$$

where $L_{\rm disk}$ is the accretion-disk luminosity. The outer radius of the BLR is subsequently defined as follows:

$$R_{\rm BLR-out} = 1.1 \times R_{\rm BLR-in}. \tag{10}$$

Similarly, the radius of the dusty torus is estimated as follows:

$$R_{\rm DT} = 2 \times 10^{19} \times \sqrt{\frac{L_{\rm disk}}{10^{46}}}\ {\rm cm}. \tag{11}$$

Panel (e) of Figure 3 displays the distributions of the outer radius of the BLR for jetted AGNs. The mean outer radius of the BLR for FSRQs, BL Lacs, BCUs, and NLS1s is $\overline{\log R_{\rm BLR-out,FSRQs}} = 17.42 \pm 0.30$, $\overline{\log R_{\rm BLR-out,BLLacs}} = 16.73 \pm 0.46$, $\overline{\log R_{\rm BLR-out,BCU}} = 17.29 \pm 0.54$, and $\overline{\log R_{\rm BLR-out,NLS1}} = 17.02 \pm 0.43$, respectively. The inner and outer radii of the BLR differ only by a factor of 1.1. Thus, we only compare the the outer radius of the BLR for jetted AGNs. Panel (f) of Figure 3 shows the distributions of the dusty torus radius for jetted AGNs. The mean radius of the dusty torus for FSRQs, BL Lacs, BCUs, and NLS1s is $\overline{\log R_{\rm DT,FSRQs}} = 19.21 \pm 0.30$, $\overline{\log R_{\rm DT,BLLacs}} = 18.51 \pm 0.46$, $\overline{\log R_{\rm DT,BCU}} = 19.07 \pm 0.54$, and $\overline{\log R_{\rm DT,NLS1}} = 18.79 \pm 0.42$, respectively.

To constrain the location of the gamma-ray emission region, we compare the distance of the emission region from the central SMBH with the physical dimensions of both the BLR and the dusty torus. The distribution of the ratio of the distance of the emission region from the central SMBH to the inner BLR radius is shown in panel (g) of Figure 3. The average values of this ratio for FSRQs, BL Lacs, BCUs, and NLS1s are $\overline{\log R/R_{\rm BLR-in,FSRQs}} = 2.13 \pm 0.91$, $\overline{\log R/R_{\rm BLR-in,BLLacs}} = 2.47 \pm 0.85$, $\overline{\log R/R_{\rm BLR-in,BCU}} = 2.46 \pm 1.00$, and $\overline{\log R/R_{\rm BLR-in,NLS1}} = 2.42 \pm 1.06$, respectively. The distribution of the ratio of the distance of the emission region from the central SMBH to the outer BLR radius is presented in panel (h) of Figure 3. The corresponding average values are $\overline{\log R/R_{\rm BLR-out,FSRQs}} = 2.09 \pm 0.91$, $\overline{\log R/R_{\rm BLR-out,BLLacs}} = 2.43 \pm 0.85$, $\overline{\log R/R_{\rm BLR-out,BCU}} = 2.42 \pm 1.00$, and $\overline{\log R/R_{\rm BLR-out,NLS1}} = 2.38 \pm 1.06$. Panel (k) of Figure 3 shows the distribution of the ratio of the distance of the emission region from the central SMBH to the dusty torus radius. The average values for FSRQs, BL Lacs, BCUs, and NLS1s are $\overline{\log R/R_{\rm DT,FSRQs}} = 0.31 \pm 0.91$, $\overline{\log R/R_{\rm DT,BLLacs}} = 0.65 \pm 0.85$, $\overline{\log R/R_{\rm DT,BCU}} = 0.64 \pm 1.00$, and $\overline{\log R/R_{\rm DT,NLS1}} = 0.59 \pm 1.06$, respectively.

From the above results, it can be inferred that the ratios of the distance of the emission region from the central SMBH to both the inner and outer radii of the BLR are $R \approx 135$–$295 R_{\rm BLR-in}$ and $R \approx 123$–$270 R_{\rm BLR-out}$, whereas the ratio of the distance of the emission region from the central SMBH to the radius of the dusty torus is $R \approx 2$–$4.5 R_{\rm DT}$. These findings indicate that the gamma-ray emission region in jetted AGNs is likely located beyond the BLR and potentially could be associated with the dusty torus. A number of studies have indicated that high-energy emission originating within the BLR is unlikely, as it would be suppressed by significant gamma-ray opacity due to pair production (e.g., M. Böttcher et al. 2009; F. Tavecchio & G. Ghisellini 2012). A.-C. Donea & R. J. Protheroe (2003) and H. T. Liu & J. M. Bai (2006) also reported that the detection of gamma-ray emission in the 10–100 GeV energy range suggests that the emission originates from a region located beyond the BLR. F. Tavecchio & G. Ghisellini (2012) investigated the origin of gamma-ray emission from the flaring blazar PKS 1222+216 and concluded that the emission likely originates beyond the BLR. Similarly, L. Costamante et al. (2018) found evidence supporting the extrinsic origin of gamma-ray emission—specifically beyond the BLR—based on an analysis of 106 broad-line Fermi blazars. Numerous studies have demonstrated that the gamma-ray emission region in blazar jets may be located within the dusty torus (e.g., F. K. Schinzel et al. 2012; M. Cerruti et al. 2013; S. G. Jorstad et al. 2013; C. D. Dermer et al. 2014; C. Casadio et al. 2015; M. Böttcher & P. Els 2016), where photons emitted by the dusty torus play a critical role in explaining the observed gamma-ray spectral features.

### *3.3. Relation between the Amplitude of Variability and the Physical Parameters*

AGNs exhibit variable emission across the entire electromagnetic spectrum, with the observed variations occurring over a wide range of timescales, from as brief as several minutes to as extended as multiple decades (e.g., C. M. Urry & P. Padovani 1995; S. J. Wagner & A. Witzel 1995; J.-H. Fan 2005; R. Falomo et al. 2014; P. Padovani et al. 2017). A variety of mechanisms have been proposed to account for the variability observed in jetted AGNs. These include processes such as the injection, acceleration, and radiative cooling of particles within the jet, which may be influenced by shock waves or turbulent activity (e.g.,

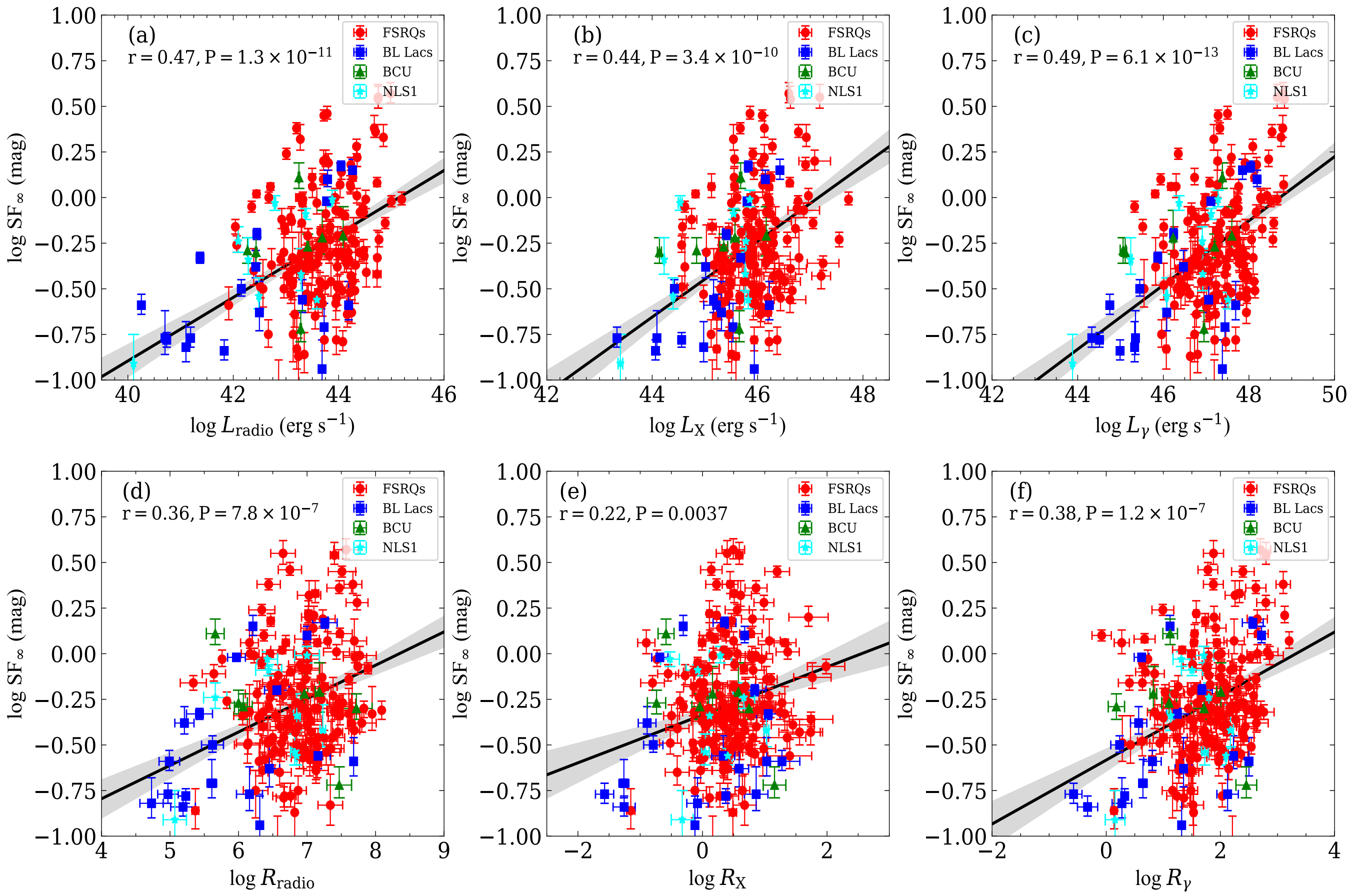

**Figure 5.** (a) The relation between the amplitude of variability and radio luminosity; (b) the relation between the amplitude of variability and 0.3–10 keV X-ray luminosity; (c) the relation between the amplitude of variability and gamma-ray luminosity; (d) the relation between the amplitude of variability and radio loudness; (e) the relation between the amplitude of variability and X-ray loudness; and (f) the relation between the amplitude of variability and gamma-ray loudness. The red dots are FSRQs. The blue dots are BL Lacs. The green dots are BCUs. The cyan dots are NLS1s. The shaded gray colored areas correspond to $1\sigma$ confidence bands. The black line is the best fit.

A. P. Marscher & W. K. Gear 1985; M. Sikora et al. 2001; G. Ghisellini et al. 2002; A. P. Marscher 2014). Additional explanations involve geometric effects or structural variations in different regions of the jet (e.g., M. Villata & C. M. Raiteri 1999; A. P. Marscher et al. 2008; C. M. Raiteri et al. 2017, 2024), phenomena related to the accretion disk (e.g., M. F. Gu et al. 2006), gravitational microlensing (e.g., D. F. Torres et al. 2003), as well as kink instabilities and magnetic reconnection events (e.g., S. G. Jorstad et al. 2022; X. Chang et al. 2024). Thus, we study the relation between the amplitude of variability and the physical parameters for jetted AGNs.

#### 3.3.1. Relation between the Amplitude of Variability and the Radio Luminosity and Radio Loudness

Some studies have confirmed a strong correlation between gamma-ray variability and emission at radio frequencies, as well as other wavelengths (e.g., S. G. Jorstad et al. 2010; J. León-Tavares et al. 2013; I. Liodakis et al. 2020; R. A. Amaya-Almazán et al. 2021). It is widely accepted that the motion of a plasma component along the jet is the primary mechanism driving most gamma-ray flare events (e.g., A. P. Marscher et al. 2010; S. G. Jorstad et al. 2013, 2017; Z. R. Weaver et al. 2019). E. Palafox et al. (2025) found a correlation between gamma-ray variability and the parsec-scale jet structure in the blazar 3C 454.3. Accordingly, we investigate the relationship between the amplitude of gamma-ray variability and jet-related physical parameters, such as the radio luminosity and gamma-ray luminosity. The relationship between the amplitude of gamma-ray variability and the 1.4 GHz radio luminosity is displayed in panel (a) of Figure 5. A statistical correlation is observed between the amplitude of gamma-ray variability and the 1.4 GHz radio luminosity for our sample (correlation coefficient $r = 0.47$, significance level $p = 1.3 \times 10^{-11}$, where $p < 0.05$ indicates a significant correlation at the 95% confidence level). The best-fit linear regression equation is as follows:

$$\log \mathrm{SF}_{\infty} = (0.17 \pm 0.025) \log L_{\mathrm{radio}} - 7.90(\pm 1.10). \quad (12)$$

Due to the small number of BL Lacs, NLS1s and BCUs, we did not perform linear regression analysis separately for each subclass in order to obtain accurate statistical results. A statistical correlation is observed between the amplitude of gamma-ray variability and radio loudness for our sample ($r = 0.36$, $p = 7.8 \times 10^{-7}$; see panel (d) of Figure 5). The best-fit linear regression equation is as follows:

$$\log \mathrm{SF}_{\infty} = (0.18 \pm 0.037) \log R_{\mathrm{radio}} - 1.52(\pm 0.25). \quad (13)$$

Radio loudness is also widely recognized as an indicator of jet activity (e.g., M. Mehdipour & E. Costantini 2019). Our findings indicate that the origin of gamma-ray variability is closely associated with jet processes. S. Rakshit & C. S. Stalin (2017) investigated the optical variability of a large sample of NLS1 and broad-line Seyfert 1 (BLS1) galaxies and found that the variability amplitude correlates with both radio loudness and radio luminosity. They suggested that jets play a significant role in the optical variability observed in radio-loud objects.

#### 3.3.2. Relation between the Amplitude of Variability and the X-Ray Luminosity and X-Ray Loudness

The X-ray emission in blazars originates primarily from strong relativistic jets pointed toward Earth, with the specific mechanism depending on the blazar type. It is generally produced by SSC (common in high-synchrotron-peaked BL Lacs) or EC (prevalent in FSRQs; e.g., J. D. Finke et al. 2008; M. Sikora et al. 2013; G. Bhatta et al. 2018; M. Böttcher & M. G. Baring 2019; I. Liodakis et al. 2025; C. M. Raiteri 2025). Because most of our samples are blazars, we therefore hypothesize that there exists an observable relationship between X-ray radiation and gamma-ray variability, with strong relativistic jets corresponding to higher X-ray power. The relationship between the amplitude of gamma-ray variability and X-ray luminosity for our sample is presented in panel (b) of Figure 5. A statistical correlation is found between the amplitude of gamma-ray variability and X-ray luminosity ($r = 0.44$, $p = 3.4 \times 10^{-10}$). The best-fit linear regression equation is as follows:

$$\log \mathrm{SF}_{\infty} = (0.21 \pm 0.03)\log L_{\mathrm{X}} - 9.86(\pm 1.43). \quad (14)$$

A statistical correlation is also observed between the amplitude of gamma-ray variability and X-ray loudness for our sample ($r = 0.22$, $p = 0.0037$; see panel (e) of Figure 5). The best-fit linear regression equation is as follows:

$$\log \mathrm{SF}_{\infty} = (0.13 \pm 0.04)\log R_{\mathrm{X}} - 0.33(\pm 0.02). \quad (15)$$

Our results may imply a compelling scenario in which variations in gamma-ray emissions from jetted AGNs are closely associated with relativistic jets. These gamma rays for jetted AGNs may be generated through SSC or EC processes. A. Shukla & K. Mannheim (2020) reported that the gamma-ray flare from the quasar 3C 279 originated from relativistic jets. Y. Chen et al. (2025) found that there is a correlation between the optical variability timescale and magnetic field of jets in jetted AGNs. They suggested that the relativistic jets drive the intrinsic variability observed in the light curves emitted by the AGN accretion disks.

#### 3.3.3. Relation between the Amplitude of Variability and the Gamma-ray Luminosity and Gamma-ray Loudness

The gamma-ray emission is believed to originate from the jet, likely from regions where particles are accelerated to relativistic energies. Accordingly, we investigate the relationship between the amplitude of variability and gamma-ray luminosity. The relationship between the amplitude of variability and gamma-ray luminosity is presented in panel (c) of Figure 5. A statistical correlation is found between the amplitude of gamma-ray variability and gamma-ray luminosity for our sample ($r = 0.49$, $p = 6.1 \times 10^{-13}$). The best-fit linear regression equation is as follows:

$$\log \mathrm{SF}_{\infty} = (0.18 \pm 0.02)\log L_{\gamma} - 8.57(\pm 1.09). \quad (16)$$

Our above results may further imply that relativistic jets play an important role in the variability of gamma rays. Essentially, the gamma-ray loudness of jetted AGNs serves as a strong indicator of their overall variability across multiple wavelengths, with stronger gamma-ray emission typically accompanied by more pronounced and rapid variations at other frequencies (e.g., B. Z. Dai et al. 2001; J. L. Richards et al. 2014). The relationship between the amplitude of variability and gamma-ray loudness is presented in panel (f) of Figure 5. A statistically significant correlation is observed between the amplitude of gamma-ray variability and gamma-ray loudness for our sample ($r = 0.38$, $p = 1.2 \times 10^{-7}$). The best-fit linear regression equation is as follows:

$$\log \mathrm{SF}_{\infty} = (0.18 \pm 0.03)\log R_{\gamma} - 0.58(\pm 0.05). \quad (17)$$

Our findings may imply that gamma-ray-loud sources tend to exhibit higher gamma-ray variability amplitudes. J. L. Richards et al. (2014) reported that the radio variability of gamma-ray-loud blazars is more pronounced than that of gamma-ray-quiet blazars.

#### 3.3.4. Relation between the Amplitude of Variability and the Synchrotron and IC Peak Frequency Luminosity

Variability can be attributed to changes in the magnetic field, particle density, and acceleration mechanisms (V. S. Paliya et al. 2021). The synchrotron peak frequency luminosity is associated with the magnetic field strength (M. Böttcher & C. D. Dermer 2010; V. S. Paliya et al. 2021). Therefore, we examine the relationship between the amplitude of variability and the synchrotron peak frequency luminosity. The relationship between the amplitude of variability and the synchrotron peak frequency luminosity is presented in panel (a) of Figure 6. A statistical correlation is observed between the amplitude of variability and the synchrotron peak frequency luminosity for our sample ($r = 0.42$, $p = 2.8 \times 10^{-9}$). The best-fit linear regression equation is as follows:

$$\log \mathrm{SF}_{\infty} = (0.19 \pm 0.03)\log L_{\mathrm{sy}} - 9.00(\pm 1.40). \quad (18)$$

M. Böttcher & C. D. Dermer (2010) reported that the luminosity of the synchrotron peak frequency is proportional to the square of the magnetic field strength, $L_{\mathrm{sy}} \propto B^2$. It was found that the optical variability of the very-high-energy gamma-ray blazar 1ES 1011+496 is primarily driven by variations in the magnetic field. Our findings indicate that the variability of gamma-ray emission is associated with the strength of the magnetic field.

A statistical correlation is also observed between the amplitude of variability and the IC peak frequency luminosity for our sample ($r = 0.47$, $p = 8.7 \times 10^{-12}$; see panel (b) of Figure 6). The best-fit linear regression equation is as follows:

$$\log \mathrm{SF}_{\infty} = (0.17 \pm 0.02)\log L_{\mathrm{IC}} - 8.00(\pm 1.10). \quad (19)$$

V. S. Paliya et al. (2021) reported that the IC peak frequency luminosity is proportional to the energy density of the external target photon field, $L_{\mathrm{IC}} \propto u_{\mathrm{ext}}$. Our findings suggest that the variability of gamma-ray emission may be associated with the energy density of the external target photon field. Several theoretical models have been proposed to explain blazar variability, which involve a transient enhancement in the energy density of the external target photon field that contributes to IC scattering. These include the synchrotron mirror model, wherein a portion of the synchrotron radiation from the jet is reflected by a nearby cloud along the jet's trajectory (M. Böttcher & C. D. Dermer 1999; M. Tavani et al. 2015) or by a stationary feature within the jet, such as in the "ring-of-fire" model (N. R. MacDonald et al. 2017). As the emission region moves past the mirror, enhanced radiative cooling is expected to reduce the peak of the electron energy distribution. This reduction, in turn, leads to a corresponding decrease in both the synchrotron and IC peak frequencies, as well as a decline in the synchrotron peak flux. If these were the sole effects at play in such a scenario, one would expect an increase in Compton dominance (CD = $L_{IC}/L_{sy}$, J. D. Finke 2013) that is correlated with increasing flux.

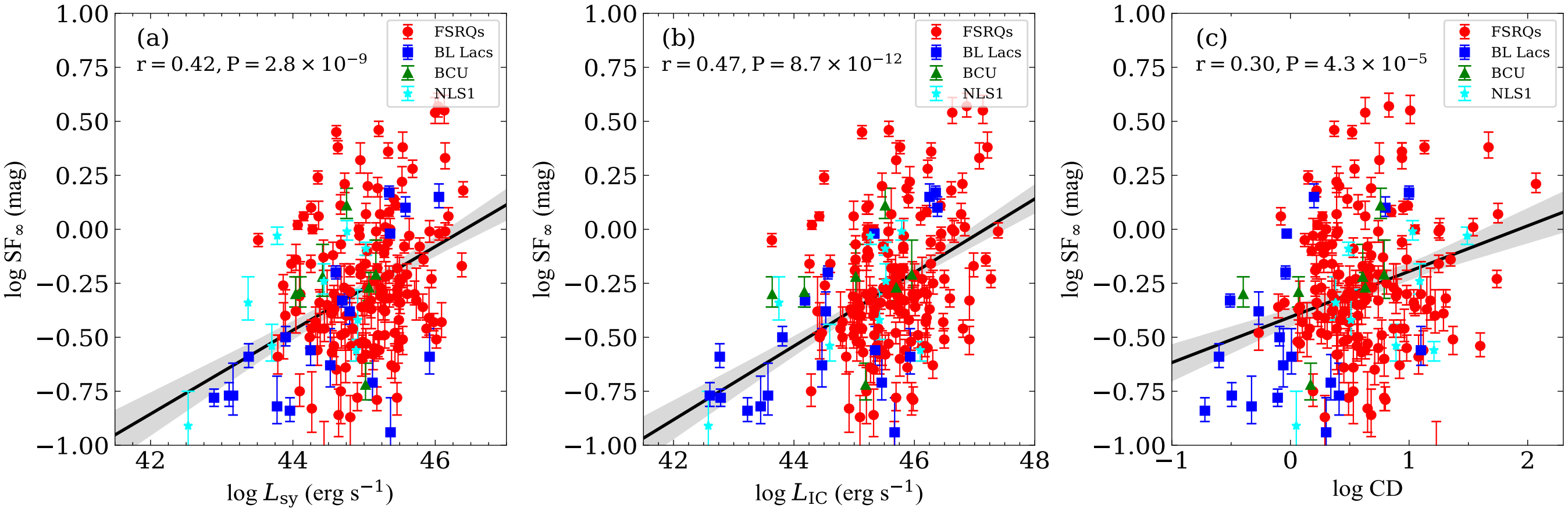

**Figure 6.** (a) The relation between the amplitude of variability and the synchrotron peak frequency luminosity; (b) the relation between the amplitude of variability and IC peak frequency luminosity; (c) the relation between the amplitude of variability and Compton dominance parameter. The red dot represents FSRQs. The blue dot represents BL Lacs. The green dot represents BCU. The cyan dot represents NLS1. Shaded gray areas correspond to 1$\sigma$ confidence bands. The black line is the best fit.

The relationship between the amplitude of variability and Compton dominance for our sample is presented in panel (c) of Figure 6. A statistical correlation is observed between the amplitude of variability and Compton dominance ($r$ = 0.30, $p = 4.3 \times 10^{-5}$). The best-fit linear regression equation is as follows:

$$\log \mathrm{SF}_{\infty} = (0.21 \pm 0.05)\log \mathrm{CD} - 0.41(\pm 0.04). \tag{20}$$

The Compton dominance parameter indicates the intensity of relativistic jets in blazars. A high Compton dominance value signals powerful jets (e.g., J. D. Finke 2013; K. Nalewajko & M. Gupta 2017; Y. Chen et al. 2023). Our results further suggest that relativistic jets enhance gamma-ray variability. A spectral hardening of the electron energy distribution, which is commonly predicted by various particle acceleration mechanisms—such as internal shocks or magnetic reconnection—results in a shift of the electron spectrum's peak toward higher energies. Assuming no other parameters are modified, this shift leads to a proportional increase in both the synchrotron and Compton peak frequencies, as well as their corresponding peak fluxes (V. S. Paliya et al. 2021). Recently, an alternative explanation was proposed by E. Sobacchi & Y. E. Lyubarsky (2019) and E. Sobacchi et al. (2021). Their hypothesis posits that stochastic particle acceleration, driven by magnetohydrodynamic turbulence, causes the most energetic particles to maintain small pitch angles relative to the magnetic field. This effect inherently results in a decrease in both the synchrotron peak frequency and its associated flux. Such a process may give rise to flaring activity through enhanced particle acceleration, leading to a shift of the peak of the electron energy distribution toward higher energies. Although this mechanism does not significantly enhance the synchrotron peak, it has the potential to substantially increase both the frequency and flux of the IC peak.

### *3.3.5. Relation between the Amplitude of Variability and Black Hole Mass and the Eddington Ratio*

Several studies have investigated the relationship between the amplitude of variability and both the black hole mass and accretion rate (e.g., M. Wold et al. 2007; B. C. Wilhite et al. 2008; B. C. Kelly et al. 2009; Y. L. Ai et al. 2010; C. L. MacLeod et al. 2010; T. Simm et al. 2016; S. Rakshit & C. S. Stalin 2017; S. Li et al. 2018; P. Sánchez-Sáez et al. 2018). We similarly examine the relationship between the amplitude of variability and these two parameters—namely, the black hole mass and accretion rate. The relationship between the amplitude of variability and black hole mass is presented in panel (a) of Figure 7. A statistical correlation is observed between the amplitude of variability and black hole mass for our sample ($r = 0.22$, $p = 0.0033$). The best-fit linear regression equation is as follows:

$$\log \mathrm{SF}_{\infty} = (0.15 \pm 0.05)\log M_{\mathrm{BH}} - 1.60(\pm 0.46). \tag{21}$$

M. Wold et al. (2007) analyzed a sample of approximately 100 quasars obtained from the Quasar Equatorial Survey Team Phase 1 variability survey. Their analysis revealed a statistically significant positive correlation between black hole mass and the amplitude of variability. B. C. Wilhite et al. (2008) reported a statistically significant positive relationship between the amplitude of variability and the mass of the central black hole. C. L. MacLeod et al. (2010) similarly observed a statistically significant positive correlation between the amplitude of optical variability and the mass of the SMBH. S. Rakshit & C. S. Stalin (2017) conducted a study based on a large sample of NLS1 and BLS1 galaxies selected from the

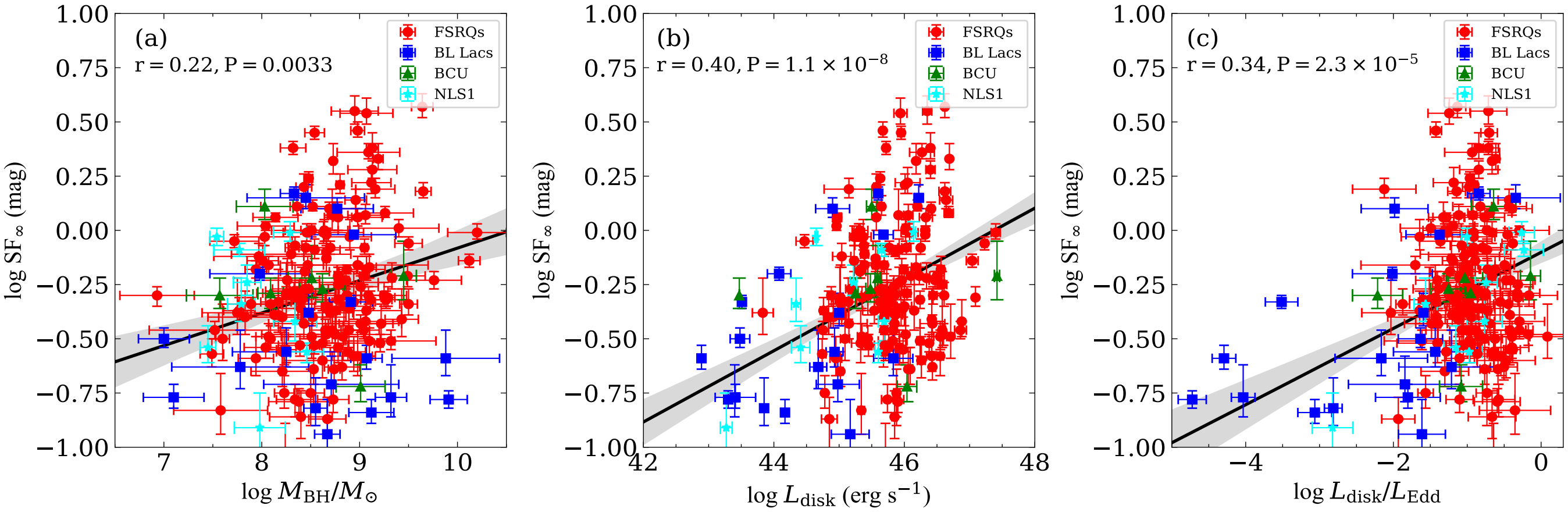

**Figure 7.** (a) The relation between the amplitude of variability and the black hole mass; (b) the relation between the amplitude of variability and the accretion-disk luminosity; and (c) the relation between the amplitude of variability and accretion rates. The red dots represent FSRQs, the blue dots represent BL Lacs, the green dots represent BCUs, and the cyan dots represents NLS1s. The shaded gray areas correspond to $1\sigma$ confidence bands. The black line is the best fit.

Catalina Real-Time Transient Survey (CRTS). Their findings indicated a significant positive correlation between the amplitude of variability and the mass of the central black hole. Our results are consistent with these previous findings.

S.-L. Li & X. Cao (2008) proposed that the positive correlation between variability amplitude and black hole mass can be explained by an accretion-disk model characterized by a mean accretion rate of $\dot{m} = 0.1$, with variations ranging from 0.1 to $0.5\dot{m}$. Thus, we study the relations between variability amplitude and accretion-disk luminosity and accretion rates for our sample. The relationship between the amplitude of variability and accretion-disk luminosity for our sample is presented in panel (b) of Figure 7. A statistical correlation is observed between the amplitude of variability and accretion-disk luminosity ($r = 0.40$, $p = 1.1 \times 10^{-8}$). The best-fit linear regression equation is as follows:

$$\log \mathrm{SF}_{\infty} = (0.16 \pm 0.03) \log L_{\mathrm{disk}} - 7.80(\pm 1.30). \quad (22)$$

There is a statistical correlation between the amplitude of variability and the Eddington ratio within our sample ($r = 0.34$, $p = 2.3 \times 10^{-5}$). The best-fitting linear regression equation is presented below:

$$\log \mathrm{SF}_{\infty} = (0.18 \pm 0.04) \log L_{\mathrm{disk}}/L_{\mathrm{Edd}} - 0.1(\pm 0.05). \quad (23)$$

Our findings suggest that the accretion disk may play a key role in driving gamma-ray variability in jetted AGNs. Variability arising from the accretion disk can trigger changes in the relativistic jet—a process likely mediated by magnetic fields, due to the dynamical coupling between the accretion disk and relativistic jet. The shock formed at the jet base may thus be modulated by such magnetic field perturbations. As these shocks propagate outward along the jet, they traverse the radiative regions, where diffusive shock acceleration acts as the dominant mechanism for energizing relativistic electrons, which in turn gives rise to the observed variability. In blazars and other beamed AGNs, this variability is further amplified by the relativistic beaming effect. It should be noted, however, that the aforementioned mechanism is not exhaustive, and additional physical processes may also contribute to the observed variability of AGNs. M. J. Graham et al. (2020) found that the extremely variable quasars with high variability amplitude tend to have high accretion rates, by using a sample of extremely variable quasars. Our results are consistent with theirs. Previous studies have reported an anticorrelation between the amplitude of variability and the Eddington ratio, based on various datasets: B. C. Wilhite et al. (2008), using Sloan Digital Sky Survey (SDSS) quasars; C. L. MacLeod et al. (2010), using SDSS Stripe 82 quasars; T. Simm et al. (2016), using XMM-COSMOS AGNs; S. Rakshit & C. S. Stalin (2017), using a large sample of NLS1 and BLS1 galaxies from CRTS; and S. Li et al. (2018), using approximately $10^5$ quasars. It is worth noting that our results are inconsistent with these earlier findings. Two potential explanations can account for this discrepancy. First, differences in sample selection may contribute to the variation in results, as our analysis is based exclusively on jetted AGNs with high accretion rates. Almost all of our sample has a high accretion rate. Jetted AGNs with higher accretion rates exhibit stronger variability in their light curves. This means that as the rate of matter falling onto the SMBH increases, the fluctuations in the AGN's variability become more pronounced. Second, the variability amplitudes were measured in different wavelength bands: our study primarily utilizes the gamma-ray band, whereas previous studies predominantly relied on optical-band data.

## 4. Conclusion

We have conducted a study of gamma-ray variability in a large sample of jetted AGNs using archival data from Fermi-LAT. This study constitutes a significant advancement over previous research in three key aspects: (1) the number of sources included in the analysis; (2) the number of data epochs used for each source; and (3) the application of random walk models to investigate the properties of gamma-ray light curves for a large sample. Our main findings are summarized as follows.

1. The mean amplitude of variability is found to be $10^{-0.26 \pm 0.32}$ for FSRQs, $10^{-0.57 \pm 0.41}$ for BL Lacs, $10^{-0.27 \pm 0.22}$ for BCUs, and $10^{-0.35 \pm 0.28}$ for NLS1s. The KS test reveals that there is a statistically significant difference in the distribution of variability amplitudes between FSRQs and BL Lacs. The mean timescale of variability is $10^{1.70 \pm 0.42}$ for FSRQs, $10^{2.07 \pm 0.37}$ for BL Lacs, $10^{2.08 \pm 0.24}$ for BCUs, and $10^{1.95 \pm 0.54}$ for NLS1s. The KS test further indicates that there is a statistically significant difference in the distribution of variability timescales between FSRQs and BL Lacs. The mean variability damping timescales of our sources are approximately 100 days. This damping timescale may imply that diffusive shock acceleration plays an important role in the variability of gamma-ray emission.

2. The mean size of the emission region in jets is $10^{18.17 \pm 0.53}$ for FSRQs, $10^{18.18 \pm 0.55}$ for BL Lacs, $10^{18.45 \pm 0.44}$ for BCUs, and $10^{18.22 \pm 0.72}$ for NLS1s. The KS test further indicates that there is no statistically significant difference in the distribution of emission-region sizes between FSRQs and BL Lacs.

3. The average values of the ratio of the distance of the emission region from the central SMBH to the inner radius of the BLR are $10^{2.13 \pm 0.91}$ for FSRQs, $10^{2.47 \pm 0.85}$ for BL Lacs, $10^{2.46 \pm 1.00}$ for BCUs, and $10^{2.42 \pm 1.06}$ for NLS1s. The average values of the ratio of the distance of the emission region from the central SMBH to the outer BLR radius are $10^{2.09 \pm 0.91}$ for FSRQs, $10^{2.43 \pm 0.85}$ for BL Lacs, $10^{2.42 \pm 1.00}$ for BCUs, and $10^{2.38 \pm 1.06}$ for NLS1s. The average values of the ratio of the distance of the emission region from the central SMBH to the radius of the dusty torus are $10^{0.31 \pm 0.91}$ for FSRQs, $10^{0.65 \pm 0.85}$ for BL Lacs, $10^{0.64 \pm 1.00}$ for BCUs, and $10^{0.59 \pm 1.06}$ for NLS1s. We find that the ratio of the distance of the emission region from the central SMBH to the dusty torus radius for our sources is $R \approx 2\text{–}4.5 R_{\rm DT}$. In contrast, the ratio of the distance of the emission region from the central SMBH to the BLR radius for our sources is $R \approx 135\text{–}295 R_{\rm BLR-in}$ and $R \approx 123\text{–}270 R_{\rm BLR-out}$. These findings indicate that the gamma-ray emission region in jetted AGNs is likely located beyond the BLR and potentially could be associated with the dusty torus.

4. There is a correlation between the amplitude of variability and the radio luminosity, radio loudness, X-ray luminosity, X-ray loudness, gamma-ray luminosity, and gamma-ray loudness in jetted AGNs. These findings suggest that gamma-ray variability may be associated with the jet activity.

5. We also find a correlation between the amplitude of variability and the synchronous peak frequency luminosity, IC peak frequency luminosity, and Compton dominance in jetted AGNs.

6. There is a correlation between the amplitude of variability and the black hole mass, accretion-disk luminosity, and Eddington ratio in jetted AGNs. These findings suggest that the accretion disk may also play a role in driving gamma-ray variability.

## Acknowledgments

Y.C. is grateful for financial support from the National Natural Science Foundation of China (No. 12203028). Y.C. is grateful for funding for the training program for talents in Xingdian, Yunnan Province (2081450001). Q.G. is supported by the National Natural Science Foundation of China (12121003, 12192220, and 12192222). This work is supported by the National Natural Science Foundation of China (11733001, U2031201, and 12433004). X.G. acknowledges the support of the National Nature Science Foundation of China (No. 12303017). This work is also supported by Anhui Provincial Natural Science Foundation project No. 2308085QA33. D.X. is supported by the NSFC (12473020), the Yunnan Province Youth Top Talent Project (YNWR-QNBJ-2020-116), and the CAS Light of West China Program.

## ORCID iDs

Yongyun Chen (陈永云) https://orcid.org/0000-0001-5895-0189
Qiusheng Gu (顾秋生) https://orcid.org/0000-0002-3890-3729
Junhui Fan (樊军辉) https://orcid.org/0000-0002-5929-0968
Dingrong Xiong (熊定荣) https://orcid.org/0000-0002-6809-9575
Xiaogu Zhong (钟晓谷) https://orcid.org/0000-0002-4626-8817
Xiaotong Guo (郭晓通) https://orcid.org/0000-0002-2338-7709
Nan Ding (丁楠) https://orcid.org/0000-0003-1028-8733